# DC Resistance Degradation of $SrTiO_3$: The Role of Virtual-Cathode Needles and Oxygen Bubbles

Ana Alvarez[1] and I-Wei Chen[1*]

[1] *Department of Material Science and Engineering, University of Pennsylvania, Philadelphia, PA 19104, USA*

## Abstract

This study of highly accelerated lifetime tests of $SrTiO_3$, a model semiconducting oxide, is motivated by the interest in reliable multilayer ceramic capacitors and resistance-switching thin-film devices. Our analytical solution to oxygen-vacancy migration under a DC voltage—the cause of resistance degradation in $SrTiO_3$—agrees with previous numerical solutions. However, all solutions fail to explain why degradation kinetics feature a very strong voltage dependence, which we attribute to the nucleation and growth of cathode-initiated fast-conducting needles. While they have no color contrast in $SrTiO_3$ single crystals and are nominally ″invisible,″ needle′s presence in DC-degraded samples—in silicone oil and in air—was unambiguously revealed by *in-situ* hot-stage photography. Observations in silicone oil and thermodynamic considerations of voltage boundary conditions further revealed a cooccurrence of copious oxygen bubbling and the onset of final accelerating degradation, suggesting sudden oxygen loss is a precursor of final failure. Remarkably, both undoped and Fe-doped $SrTiO_3$ can emit electroluminescence at higher current densities, thus providing a vivid indicator of resistance degradation and a metal-to-insulator resistance transition during cooling. The implications of these findings to thin ceramic and thin film $SrTiO_3$ devices are discussed, along with connections to similar findings in likewise degraded fast-ion yttria-stabilized zirconia.

*Corresponding Author, email: iweichen@seas.upenn.edu

## INTRODUCTION

Like most insulators, acceptor-doped perovskite titanates such as $SrTiO_3$ (STO) may breakdown under a DC voltage ($\Delta\phi$) or field ($E$). Dielectric breakdown has long been studied by highly accelerated lifetime tests (HALTs),[1-4] which continues today with new attention to multilayer ceramic capacitors (MLCCs)[5] and resistance random memory access memories (ReRAMs)[6-8]. Collected from bulk samples, HALT data on voltage-induced resistance ($R$) degradation are used to infer the lifetime ($t^{f}$) and reliability of thin ceramic and nano film devices.[9-17] HALT is also relevant to flash sintering[18-20] of titanates,[21-24] for both phenomena feature an abrupt, huge rise in current density ($J$) and Joule heating. Therefore, a better understanding of HALT in STO will benefit several scientific and technological fields.

The stage of current consensus was set by Waser *et al.*[2-4] almost 30 years ago: Titanates degrade because of electromigration of quenched-in oxygen vacancies ($V_O^{\bullet\bullet}$), which turns a test piece from a sluggish ionic conductor into a highly conductive *p-n* electronic junction. The difficult mathematical problem (see **Governing equation and scaling law**) was also tackled by them and later researchers numerically.[4,11-12,14-15] Surprisingly, the solutions revealed a rather weak field dependence of kinetics, *e.g.*, $t^f \propto E^{-0.61}$,[11] which is not supported by experimental data (reviewed in annotated preprint[25-A1]) including Waser's own.[2-3] Clearly, something is missing in current understanding.

This study will (a) present an analytical solution that relates $t^f$ to $E$ and other test and sample parameters, and (b) reconcile the discrepancy in the $E$ dependence by presenting new observations of needle-like short-circuit paths, whose nucleation and growth can cause a strong $E$ dependence. To reveal needle-growth in real time, we tested undoped and Fe-doped transparent single crystals in air and silicone oil, with extraneous electrocoloration avoided in the latter because it suppresses surface reoxygenation by environmental oxygen. Noting similar DC degradation of $Y_2O_3$ (8 mol%)-stabilized zirconia (8YSZ), which is a fast $V_O^{\bullet\bullet}$ conductor easily losing oxygen during HALT,[26] we discovered degraded STO also releases oxygen bubbles profusely, and like 8YSZ, becomes a metal and may further transition to an insulator during cooling. Lastly, we discovered STO can emit current-dependent electroluminescence, which sends a visual signal for resistance degradation and metal-insulator transition.

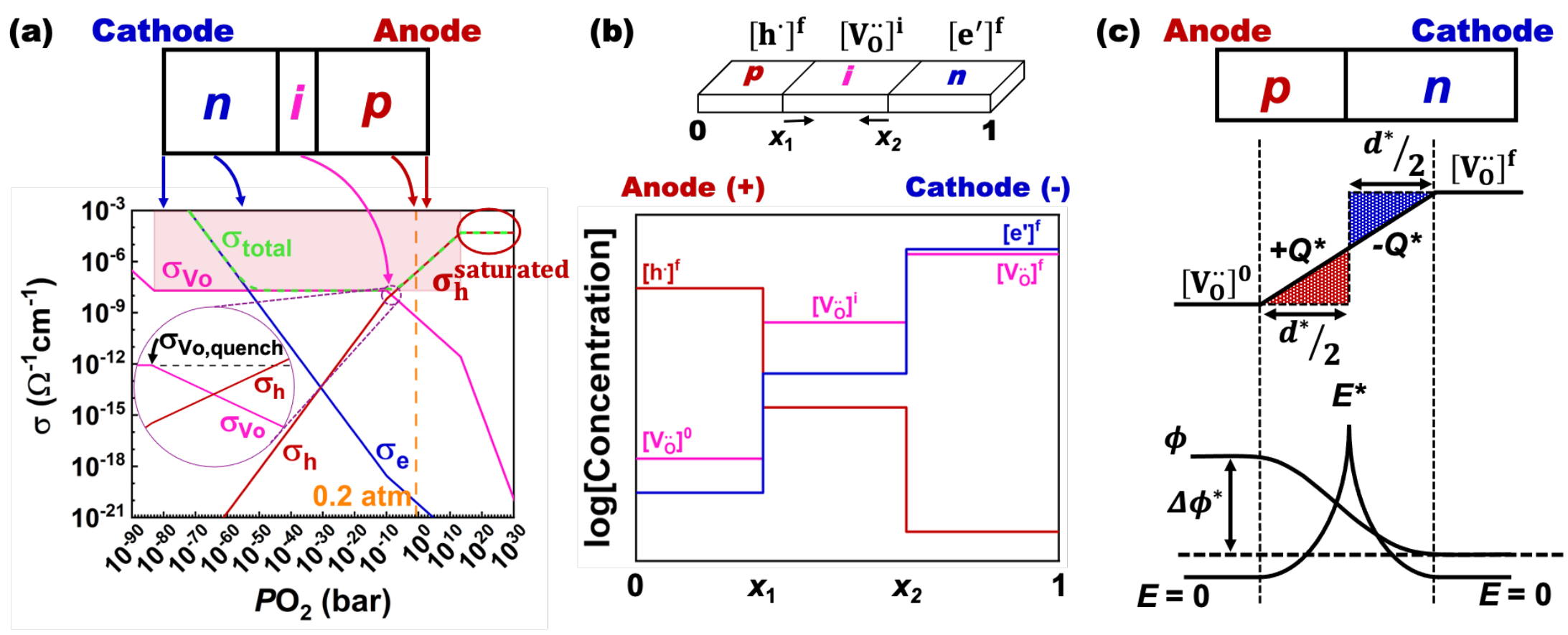

**Figure 1** (a) Calculated (see **Appendix**) $\sigma_{V_O}$, $\sigma_e$ and $\sigma_h$ plus their sum vs $PO_2$ at 210°C with $5.58\times10^{18}$ cm$^{-3}$ Fe concentration. Top: Sample schematic with different conductivity regions. $\left[Fe'_{Ti}\right]\approx 2\left[V_O^{\bullet\bullet}\right]$ in shaded "extrinsic" region and $\left[Fe'_{Ti}\right]\approx\left[h^{\bullet}\right]$ in oval-circled "saturated" region. Quenched-in $\left[V_O^{\bullet\bullet}\right]$ as black broken line sets initial vacancy conductivity and internal $PO_2$ at arrowed intercept in circled inset. (b) Sample's *p-i* interface at $x_1$ moves to right and *i-n* interface at $x_2$ moves to left, both driven by electromigration of $V_O^{\bullet\bullet}$. Box profiles assumed for $\left[V_O^{\bullet\bullet}\right]$, $\left[e'\right]$ and $\left[h^{\bullet}\right]$. (c) When $x_1 = x_2$, *i*-region vanishes, (top) *p-n* junction forms, and (middle) redistributed $\left[V_O^{\bullet\bullet}\right]$ from charge-neutral box profile to straight-line profile causes space charge $\pm Q^*$ (shaded in colors) across junction thickness $d^*$, building up (bottom) junction electric field $E^*$ and potential $\Delta\phi^*$.

## SYNOPSIS

### Defect chemistry and conductivity bifurcation

At the temperature of interest, cations in STO are immobile, and $V_O^{\bullet\bullet}$, though mobile, is sluggish enough to be quenched from the annealing temperature to the test temperature. In an ambient environment, quenched-in $V_O^{\bullet\bullet}$ dominates the conductivity of quenched STO and local $PO_2$, thus setting the local concentrations of other defects as described in **Appendix**. Donor-doped STO is highly conductive and cannot be degraded. So, we studied crystals of an acceptor (Fe)-doped STO with increased $V_O^{\bullet\bullet}$ concentration ($[V_O^{\bullet\bullet}]$) and an "undoped" STO with some trace Fe impurity. At the test temperature, both quenched crystals started almost colorless and acquired a brown/black color from electro-oxidized $Fe^{4+}$ during testing.

As in all redox-active oxides, electronic conductivity crossovers from electron conductivity $\sigma_e$ dominance (*n*-type) to hole conductivity $\sigma_h$ dominance (*p*-type) as $PO_2$ increases in **Figure 1a**. (More plots in annotated preprint.[25-A2]) At mid-$PO_2$, $[V_O^{\bullet\bullet}]$ is limited by Fe-doping, thus a flat $V_O^{\bullet\bullet}$ conductivity ($\sigma_{V_O}$) dominates ("*i*-type"). An STO pushed beyond such a range is an *n*-type or *p*-type semiconductor.[11] This can be achieved by redistributing quenched-in $V_O^{\bullet\bullet}$ (dotted black line in the circled inset of **Figure 1a**), via electromigration, thus altering the local $PO_2$ in different parts of the sample. HALT can thus create, in **Figure 1a-b**, (i) a highly reduced, $V_O^{\bullet\bullet}$-accumulating, $Fe^{3+}$-rich *n*-region near the cathode, and (ii) a highly oxidized, $V_O^{\bullet\bullet}$-depleting, $Fe^{4+}$-rich (black/brown) *p*-region near the anode. Given the boat shape of total conductivity $\sigma_{total}$ in **Figure 1a** and the initial state at the bottom of $\sigma_{total}$, any carrier-segregating STO will experience a conductivity bifurcation that increases the overall conductivity. As $[V_O^{\bullet\bullet}]$-mediated carrier segregation and conductivity bifurcation becomes more severe, more resistance degradation follows. This is seen at a higher $E$, but also at a lower temperature where the boat shape looks more extreme.[25-A2] This is the physics of DC resistance degradation in STO.

Not only severity but also lengths (and kinetics) of different regions in **Figure 1a-b** depend on $E$, temperature, sample length, time and electrode's ability to block oxygen: ingress at cathode and egress at anode. A huge increase in $J$ first comes when *n*- and *p*-regions impinge each other (**Figure 1c**), followed by the gradual establishment of a purely electronic steady-state $J^{ss}$ if oxygen exchange is completely blocked. Such state has no oxygen current, so $\sigma_{V_O}$ no longer matters. Therefore, the *p-n* junction must pass through the V-shape bottom in **Figure 1a**, a minimum in total electronic conductivity where $\sigma_e$ intersects $\sigma_h$. On the other hand, if sufficient oxygen egress (loss) from the anode occurs, then no steady state is obtained and both the junction and its conductivity minimum can be eliminated: The *p*-region is gone, the *n*-region remains, and $J$ increases continuously. Lastly, **Figure 1a** dictates that a highly reduced *n*-region is much more conductive than a highly oxidized *p*-region, the latter's $\sigma_h$ saturating at $5\times10^{-5}\ \Omega^{-1}\text{cm}^{-1}$ (oval-circled in **Figure 1a**).

**Governing equation and scaling law**

Since $V_O^{\bullet\bullet}$ electromigration sets the local $PO_2$ and defect concentrations, one must first solve the $\left[V_O^{\bullet\bullet}\right]$ distribution, under an electric potential $\phi$, using the following continuity equation

$$\partial\left[V_O^{\bullet\bullet}\right]/\partial t = \nabla \bullet \left(\left[V_O^{\bullet\bullet}\right] M_{V_O} \nabla\left(kT\ln\left[V_O^{\bullet\bullet}\right] + z_{V_O} e\phi\right)\right) \qquad (1)$$

Here, both an entropic force $-\nabla\left(kT\ln\left[V_O^{\bullet\bullet}\right]\right)$ and an electric force $-\nabla\left(z_{V_O} e\phi\right)$ drive $V_O^{\bullet\bullet}$ electromigration, with $M_{V_O}$ and $z_{V_O}e = 2e$ the mobility and charge of $V_O^{\bullet\bullet}$, respectively, $k$ the Boltzmann constant and $T$ the absolute temperature. For a sample of length $L$ and a characteristic time, say failure time $t^{\mathrm{f}}$, the solution of Eq. (3) obeys a scaling law, $M_{\mathrm{V_O}} t^{\mathrm{f}} \propto L^2$. However, since Eq. (1) is nonlinear, changing the quenched-in $\left[\mathrm{V_O^{\bullet\bullet}}\right]$ or the electric potential $\phi$ by a factor $\lambda$ does not change $t^{\mathrm{f}}/L^2$ by $\lambda$ or $1/\lambda$.

## ANALYTICAL SOLUTION

### Degradation kinetics and lifetime

To make Eq. (1) solvable, we linearize it by using a box-shaped $\left[\mathrm{V_O^{\bullet\bullet}}\right]$ in **Figure 1b**. Here the degrading slab of length $L$ is expressed in a normalized coordinate $x$, its $p$-region with a hole concentration $\left[\mathrm{h^\bullet}\right]^{\mathrm{f}}$ from $x=0$ to $x_1$, $i$-region with $\left[\mathrm{V_O^{\bullet\bullet}}\right]=\left[\mathrm{V_O^{\bullet\bullet}}\right]^{\mathrm{i}}$ from $x_1$ to $x_2$, and $n$-region with $\left[\mathrm{e'}\right]^{\mathrm{f}}$ from $x_2$ to 1. The above concentrations $\left[\mathrm{d}\right]$ (d = $\mathrm{h^\bullet}$, $\mathrm{e'}$, or $\mathrm{V_O^{\bullet\bullet}}$) are for the conduction-dominant defects only, and their superscripts (i or f) denote that they are the same as in the initial (i) or final (f) state solution. In solving Eq. (1), it suffices to treat $\left[\mathrm{V_O^{\bullet\bullet}}\right]=\left[\mathrm{e'}\right]=0$ in the $p$-type region, $\left[\mathrm{h^\bullet}\right]=\left[\mathrm{e'}\right]=0$ in the $i$-type region, $\left[\mathrm{h^\bullet}\right]=0$ in the $n$-type region, and $\left[\mathrm{V_O^{\bullet\bullet}}\right]=\left[\mathrm{V_O^{\bullet\bullet}}\right]^{\mathrm{f}}>\left[\mathrm{V_O^{\bullet\bullet}}\right]^{\mathrm{i}}$ in the $n$-region where $\mathrm{V_O^{\bullet\bullet}}$ is accumulated. Later, we will lump all nonlinearity back to $\left[\mathrm{V_O^{\bullet\bullet}}\right]^{\mathrm{f}}$ when comparing with numerical solutions.

Using similar procedures in our 8YSZ paper,[26] we solve the moving interface problem as follows (details in annotated preprint[25-A3]). First, in each region, the conductivity $\sigma_{\mathrm{d}}$ from the respective dominant defect is $\sigma_{\mathrm{d}}=\left(z_{\mathrm{d}}e\right)^2 M_{\mathrm{d}}\left[\mathrm{d}\right]$, where $M_{\mathrm{d}}=D_{\mathrm{d}}/kT$, with $D_{\mathrm{d}}$ the defect diffusivity and $z_{\mathrm{d}}e$ the defect charge. This information is enough to determine (i) the respective resistance of the three regions, which form a set of serial resistors, and (ii) the potential drop in each region subject to the boundary potential: $\phi_A$ at the anode and $\phi_C = \phi_A - \Delta\phi$ at the cathode. Second, using the quenched-in population $\left[V_O^{\bullet\bullet}\right]^i$ and the potential drop in the *i*-region, we obtain the $V_O^{\bullet\bullet}$ flux *j*. Third, applying the continuity equation for $\left[V_O^{\bullet\bullet}\right]$ at $x_1$ and $x_2$ to conserve $V_O^{\bullet\bullet}$ across the concentration discontinuities $\Delta\left[V_O^{\bullet\bullet}\right]_{1,2}$, we obtain $\left|dx_{1,2}/dt\right| = j/L\Delta\left[V_O^{\bullet\bullet}\right]_{1,2}$, the speeds at which the two interfaces move toward each other. Fourth, integration gives the following solution for $x_1$ and $x_2$ if the resistance of the *n*-region, which has a very high $\sigma_e$, is ignored.

$$x_1 = \frac{x^f}{1-\gamma}\left(1-\sqrt{1-\frac{t}{t^f}+\gamma^2\left(\frac{t}{t^f}\right)}\right) \tag{2}$$

$$x_2 = 1-\left(\frac{\left[V_O^{\bullet\bullet}\right]^i}{\left[V_O^{\bullet\bullet}\right]^f-\left[V_O^{\bullet\bullet}\right]^i}\right)x_1 \tag{3}$$

Here, $x = x^f$ and $t = t^f$ are, respectively, where and when the two interfaces meet,

$$x^f = \frac{\left[V_O^{\bullet\bullet}\right]^f-\left[V_O^{\bullet\bullet}\right]^i}{\left[V_O^{\bullet\bullet}\right]^f} \tag{4}$$

$$t^f = x^f\left(1+\gamma\right)\left(\frac{kT}{z_{V_O}e\Delta\phi}\right)\left(\frac{L^2}{2D_{V_O}}\right) \tag{5}$$

and $\gamma$ is a constant,

$$\gamma = \left(\frac{z_{V_O}^2 M_{V_O}\left[V_O^{\bullet\bullet}\right]^i}{z_h^2 M_h\left[h^\bullet\right]^f}\right)x^f \tag{6}$$

From the above, we obtain $\left[V_O^{\bullet\bullet}\right]^i = \left(1-x^f\right)\left[V_O^{\bullet\bullet}\right]^f$ and confirm $\left[V_O^{\bullet\bullet}\right]$ conservation. The current density $J$ being the same throughout the slab

$$J = \frac{\sigma_{V_O}\Delta\phi}{L\sqrt{1-\frac{t}{t^f}+\gamma^2\left(\frac{t}{t^f}\right)}} \tag{7}$$

has two limiting values

$$J\big|_{t=0} = \frac{\sigma_{V_O}\Delta\phi}{L} \tag{8a}$$

$$J\big|_{t=t^f} = \frac{\sigma_{V_O}\Delta\phi}{\gamma L} = \frac{\sigma_h\Delta\phi}{x^f L} \tag{8b}$$

with

$$\sigma_{V_O} = z_{V_O}^2 e^2 \left[V_O^{\bullet\bullet}\right]^i M_{V_O} \tag{9}$$

$$\sigma_h = z_h^2 e^2 \left[h^\bullet\right]^i M_h \tag{10}$$

It also follows

$$\gamma = J\big|_{t=0} \big/ J\big|_{t=t^f} \tag{11}$$

which completes the solution.

Defining failure in HALT as $J\left(t^{\text{failure}}\right) = 10J\big|_{t=0}$,[2] we find $t^{\text{failure}}$ almost indistinguishable from $t^f$ (see **Figure 2a**). In the highly oxidized *p*-region in **Figure 1a**, $\sigma_h$ saturates at a value ~2500× higher than $\sigma_{V_O}$ in the *i*-region. Therefore, in a typical HALT, $\gamma = J\big|_{t=0} \big/ J\big|_{t=t^f}$ should range from $10^{-2}$ to $10^{-4}$, which is consistent with the HALT data. In such limit, $\Delta\phi$ becomes entirely spent in the *i*-region and the $x_1 - t$ plot becomes a parabola, which is essentially what Eq. (2) is.

### Steady-state *p-n* junction

At the steady state, a *p-n* junction has no $V_O^{\bullet\bullet}$ flow. It must contain the V-shape minimum $\left(\sigma_e + \sigma_h\right)_{\min}$ and, across it, a space charge region like the one drawn in

**Figure 1c**. The polarity of the region is opposite to that of the depletion layer in a semiconductor *p-n* junction because its charge arises from $V_O^{\bullet\bullet}$ redistribution instead of electron and hole depletion. As our solution assumes charge neutrality and is maintained by the original box profile, a redistributed $[V_O^{\bullet\bullet}]$—the straight-line one in **Figure 1c**—will create a space charge layer with the following consequences.

(a) Without any $V_O^{\bullet\bullet}$ flow, $V_O^{\bullet\bullet}$ must have a constant electrochemical potential, giving $kT\ln\left([V_O^{\bullet\bullet}]^f/[V_O^{\bullet\bullet}]^0\right)=z_i e\Delta\phi^*$ where $\Delta\phi^*$ is the junction potential.

(b) The space charge density from $V_O^{\bullet\bullet}$ redistribution is shown as colored triangles. For $[V_O^{\bullet\bullet}]^0 \ll [V_O^{\bullet\bullet}]^f$, it peaks at $\pm z_i e[V_O^{\bullet\bullet}]^f/2$ at the junction center. Integration over the half thickness $d^*/2$ then gives the areal space-charge-density $Q^*=z_i e d^*[V_O^{\bullet\bullet}]^f/8$. Next, from the Gauss Law, the junction-center peak field is obtained: $E^*=Q^*/\varepsilon=z_i e d^*[V_O^{\bullet\bullet}]^f/8\varepsilon$ where $\varepsilon$ is the permittivity of STO.

(c) Integration of the parabola-shaped *E* across $d^*$ gives the junction potential $\Delta\phi^*=z_i e d^{*2}[V_O^{\bullet\bullet}]^f/24\varepsilon$, which must be consistent with (a) thus setting $d^*$.

(d) At the junction center that coincides with the conductivity minimum, $E^*$ itself can drive an electronic $J^*=(\sigma_e+\sigma_h)_{min}E^*$.

From the above, we obtain

$$E^*=\sqrt{\left(\frac{3kT}{8\varepsilon}\right)[V_O^{\bullet\bullet}]^f\ln\left(\frac{[V_O^{\bullet\bullet}]^f}{[V_O^{\bullet\bullet}]^0}\right)} \tag{12a}$$

$$d^*=\sqrt{\left(\frac{24\varepsilon kT}{z_i^2e^2[V_O^{\bullet\bullet}]^f}\right)\ln\left(\frac{[V_O^{\bullet\bullet}]^f}{[V_O^{\bullet\bullet}]^0}\right)} \tag{12b}$$

$$\Delta\phi^* = E^* d^* / 3 \tag{12c}$$

$$J^* = (\sigma_e + \sigma_h)_{min} \sqrt{\left(\frac{3kT}{8\varepsilon}\right)[V_O^{\bullet\bullet}]^f \ln\left(\frac{[V_O^{\bullet\bullet}]^f}{[V_O^{\bullet\bullet}]^0}\right)} \tag{12d}$$

Here, $d^*$ is essentially the Debye length with screening achieved by $V_O^{\bullet\bullet}$ redistribution.

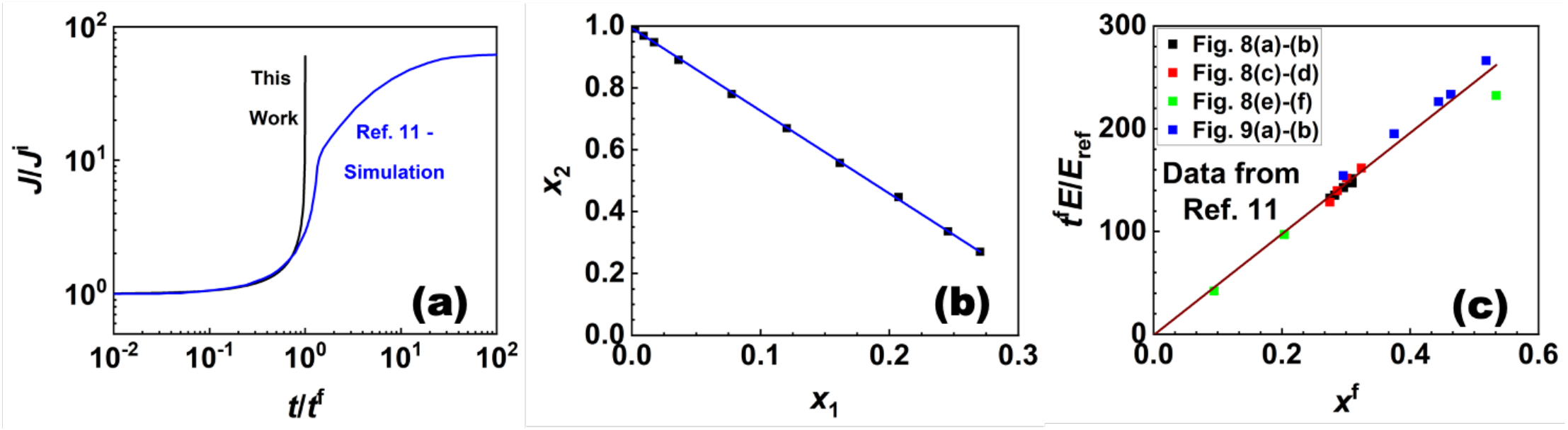

**Figure 2** (a) Analytical and numerical (from Fig. 5b of Ref. 11) solutions of $J/J^i$ vs $t/t^f$. (b) $x_2$ vs $x_1$ using data in Fig. 6 from Ref. 11. (c) $t^f E/E_{ref}$ vs $x^f$ with $E_{ref}$= 0.8 kV/cm using data in Fig. 8-9 from Ref. 11.

**Analytical solution agrees with numerical solutions**

We made the solution possible by assuming a set of box profiles for defect concentrations and disregarding their possible $E$-dependences, which allows $V_O^{\bullet\bullet}$ conservation alone to dictate interface velocities. A direct outcome of this assumption is Eq. (3): The trajectory of ($x_1, x_2$) must fall on a straight line, which is verified in **Figure 2b** by the numerical solutions of Wang *et al.* (their Fig. 6).[11] It also lends support to the assumption that there is no change in $[V_O^{\bullet\bullet}]^f$ despite extensive electromigration and $x_1/x_2$ movement. This is understandable because, as shown in **Appendix**, a flat $\left[V_O^{\bullet\bullet}\right]$ in **Figure 1a** is set by the dopant ($2\left[V_O^{\bullet\bullet}\right] = \text{Fe}$) concentration. Moreover, even with a large $\Delta\phi$, pushing $\left[V_O^{\bullet\bullet}\right]$ any higher is very difficult since, in a highly reduced state with a very high $\sigma_e$, there is little $E$ to drive further $V_O^{\bullet\bullet}$ accumulation.

Our solution also assumes $E$ is constant within each region. In contrast, all the previous work[4,11,14] emphasized the importance of sharp $E$-field-spikes at $x_1$ and $x_2$, which we disagree with. The spikes arise to overcome the entropic driving force on $V_O^{\bullet\bullet}$. For a concentration jump ($c$) of 10 times, the entropic driving force $kT\ln(c)$ at 573K for a divalent $V_O^{\bullet\bullet}$ can be countered by a potential drop of 0.055 V. So, a mere potential drop of 0.5 V at $x_1$ or $x_2$ is enough to work against $c = 10^9$, which surely exceeds $c < 10^5$, the $\left[V_O^{\bullet\bullet}\right]$ range over the entire sample length reported in all the numerical solutions![11-12,14-15] Therefore, for a typical HALT using $\Delta\phi \sim 100$ V, the potential drops at the two spikes at $x_1$ and $x_2$ have no material effect on the overall $E$ and $\phi$.

Our predicted $J(t)$ in **Figure 2a** has a similar initial shape as the numerical solution of Wang *et al.*[11] Our $J|_{t=t^f}$ is controlled by $\sigma_h$, and $J|_{t=0}$ by $\sigma_{V_O}$, at a ratio set by $\sigma_h^{\text{saturated}} / \sigma_{V_O}$, which is ~2500 in **Figure 1a**. Since Wang *et al.*'s[11] solution shows a final $J$ less than 100× of $J|_{t=0}$, we believe it is not saturated and may not be the steady state (see **Steady state possible in HALT?**).

As mentioned before, we hope including a weak $\left[V_O^{\bullet\bullet}\right]^f \propto E^\alpha$ dependence will allow our solution to capture all the essential nonlinear features of the numerical solutions. To check this, we start with Wang *et al.*'s[11] Fig. 9b, which gives $\left[V_O^{\bullet\bullet}\right]^f \propto E^{0.25}$, a weak dependence indeed. Conservation of $V_O^{\bullet\bullet}$ in its box profile thus requires $1-x^{\mathrm{f}}$ to gradually shrink with $E$, so we predict $x^{\mathrm{f}}=\left(\left[\mathrm{V}_{\mathrm{O}}^{\bullet\bullet}\right]^{\mathrm{f}}-\left[\mathrm{V}_{\mathrm{O}}^{\bullet\bullet}\right]^{\mathrm{i}}\right)/\left[\mathrm{V}_{\mathrm{O}}^{\bullet\bullet}\right]^{\mathrm{f}}$ has a weak $E$-dependence, say $x^{\mathrm{f}}\propto E^{\beta}$, which also gives $t^{\mathrm{f}}\propto x^{\mathrm{f}}L^2/\Delta\phi\propto E^{-(1-\beta)}$. This explains Wang *et al.*'s[11] (inset of their Fig. 9a) $t^{\mathrm{f}}\propto E^{-0.61}$ or $\beta=0.39$, and their $x^{\mathrm{f}}\propto E^{0.34}$ or $\beta=0.34$ (inset of Fig. 9b.) Therefore, admitting a weak $E$-dependence to $\left[\mathrm{V}_{\mathrm{O}}^{\bullet\bullet}\right]^{\mathrm{f}}$, *a posteriori*, suffices to embody in our solution the essential element required to reproduce all the nonlinear $E$-dependences in numerical solutions.

Our solution further predicts the following. First, $1/t^{\mathrm{f}}$ should have the same activation energy of $D_{\mathrm{V_O}}$, which is 0.86 eV. Wang *et al.*[11] found an activation energy of 0.84 eV, the small discrepancy coming from a slight decrease in $x^{\mathrm{f}}$ at increasing temperature. Second, since $t^{\mathrm{f}}\propto x^{\mathrm{f}}=\left(\left[\mathrm{V}_{\mathrm{O}}^{\bullet\bullet}\right]^{\mathrm{f}}-\left[\mathrm{V}_{\mathrm{O}}^{\bullet\bullet}\right]^{\mathrm{i}}\right)/\left[\mathrm{V}_{\mathrm{O}}^{\bullet\bullet}\right]^{\mathrm{f}}$, an increasing $\left[\mathrm{V}_{\mathrm{O}}^{\bullet\bullet}\right]^{\mathrm{i}}$– achievable by increasing the annealing temperature or decreasing the annealing $P\mathrm{O}_2$– will cause both $x^{\mathrm{f}}$ and $t^{\mathrm{f}}$ to decrease, and vice versa. Indeed, Wang *et al.*'s[11] Fig. 8e-f show increasing $x^{\mathrm{f}}$ and $t^{\mathrm{f}}$ with increasing annealing $P\mathrm{O}_2$. Likewise, their Fig. 8c-d show decreasing $x^{\mathrm{f}}$ and $t^{\mathrm{f}}$ with increasing annealing temperature, albeit only slightly because $\left[\mathrm{V}_{\mathrm{O}}^{\bullet\bullet}\right]^{\mathrm{i}}$ is already quite large at the pressure ($2\times10^{-5}$ bar) used. Lastly, although Wang *et al.*[11] did not test the predicted $t^{\mathrm{f}}\propto L^2$ dependence, it was implied in their work and also confirmed in the experimental study of Waser *et al.*[2]

Finally, $t^{\mathrm{f}}\propto x^{\mathrm{f}}L^2/\Delta\phi$ suggests $Et^{\mathrm{f}}$ plotted against $x^{\mathrm{f}}$ should follow a straight line. Wang *et al.*[11] presented four sets of calculated $x^{\mathrm{f}}$ and $t^{\mathrm{f}}$, each set covering a range of dopant concentrations, annealing temperatures, annealing $P\mathrm{O}_2$, and $E$, respectively. As predicted, they all fall on the same straight line in **Figure 2c**, again providing very strong evidence that our solution embodies the correct form of $E$, $L$, $x^{\mathrm{f}}$ and $t^{\mathrm{f}}$.

Junction characteristics are next estimated using numerical solutions. Reading Wang *et al.*'s[11] $\left[\mathrm{V}_{\mathrm{O}}^{\bullet\bullet}\right]^{\mathrm{f}}$ and $\left[\mathrm{V}_{\mathrm{O}}^{\bullet\bullet}\right]^{0}$ from their Fig. 8(a-f) for three sets of studies on dopant concentrations, annealing temperatures and annealing $P\mathrm{O}_2$, we estimate $d^*$ to vary from 56 to 142 nm compared to the Debye length of STO ranging from 30 to 200 nm in the literature;[27] $\Delta\phi^*$ from 0.095 V to 0.212 V compared to our earlier estimates of spike potentials; and $E^*$ from 19.9 kV/cm to 113 kV/cm compared to the nominal field of 0.8 kV/cm used by Wang *et al.*[11], thus confirming an $E$-spike does exist despite junction's miniscule $\Delta\phi^*$. Lastly, although $J^*$ is expected to be much smaller than Wang *et al.*'s[11] $J^{\mathrm{ss}}$, they are proportional to each other as shown in the annotated preprint.[25-A4]

**Two conundrums**

With our analytic solution of $\left[\mathrm{V}_{\mathrm{O}}^{\bullet\bullet}\right]$ transitions, $x^{\mathrm{f}}$ and $t^{\mathrm{f}}$ (including their temperature, field, composition and heat-treatment dependence) in excellent agreement with past numerical solutions, and with additional insight afforded by $E^*$, $d^*$ and $\Delta\phi^*$, it is now evident that there are two conundrums concerning the discrepancy between the consensus picture of $\mathrm{V}_{\mathrm{O}}^{\bullet\bullet}$ electromigration and the experimental data of resistance degradation.[25-A1]

(1) Why are experimental lifetimes in HALT so sensitive to $E$ or $\Delta\phi$, typically following $E^{-n}$, where $n > 2$ and even reaching 15.6?[25-A1] This is totally inexplicable in the consensus picture: Because the migration distance $\left(Lx^{\mathrm{f}}\right)$ increases with $E$, following $E^{\beta}$, time to failure should be longer than what the linear theory predicts, i.e., it should be $t^{\mathrm{f}} \propto E^{-1+\beta}$.

(2) Why are the "steady-state" $\left[\mathrm{V}_{\mathrm{O}}^{\bullet\bullet}\right]$ profiles in past numerical solutions claiming full degradation and zero $\mathrm{V}_{\mathrm{O}}^{\bullet\bullet}$ flow only spanning such a narrow $\left[\mathrm{V}_{\mathrm{O}}^{\bullet\bullet}\right]$ range—less than $10^5$?[4,11] As we illustrated earlier, a $10^5$ range of $\left[\mathrm{V}_{\mathrm{O}}^{\bullet\bullet}\right]$ can at most withstand 0.28 V, which is miniscule compared to the typical 100-1000 V used for HALT of macroscopic samples, 70-1000 V for HALT of MLCC, and 8-50 V for testing/forming thin film devices.[25-A1]

## EXPERIMENTS

### Experimental procedure

To resolve the above conundrums we conducted over 90 HALT at up to 200 V from 200°C to 490°C using undoped (MTI Corp., total impurity level < 25 ppm wt. with 1 ppm wt. Fe) and 0.05 wt.% Fe-doped (MSE Supplies) STO single crystals following the procedure in our 8YSZ study.[26] Test bars of 5 mm long, 1.6 mm wide and 0.5 mm thick with pure Pt electrodes, made from a paste (Fuel Cell Materials) applied to two opposite ends on the same face, were annealed at 1000°C (2 h). To steer electrical field along the body diagonal, some bars had electrodes placed at opposite corners on opposite faces. Most tests were conducted on a hot plate, in air, but some were in silicone oil to minimize oxygen ingress and to observe oxygen egress. Color-band development, electroluminescence and oxygen bubble evolution were captured in real time by a video camera aided by additional lenses, and only *in-situ* photos are shown below.

## Generic degradation curve

Like 8YSZ, semi-log resistance-time $R(t)$ curves of degrading STO crystals exhibit a three-stage behavior[26] depicted in **Figure 3a**. Sometimes extremely brief, Stage I is concave downward and followed by a concave upward Stage II. At lower temperatures, HALT often ends here with a resistance that seems to have stopped degrading. This is not the steady state, though, and as will be demonstrated later (see **Figure 9**), accelerated degradation resumes after some incubation time. At higher temperatures, Stage II is compressed, and degradation soon reaccelerates and enters Stage III, which is concave downward. It finally ends in rapid sample fracture or electrode detachment, which can be avoided by setting a current compliance, typically 20 mA or less.

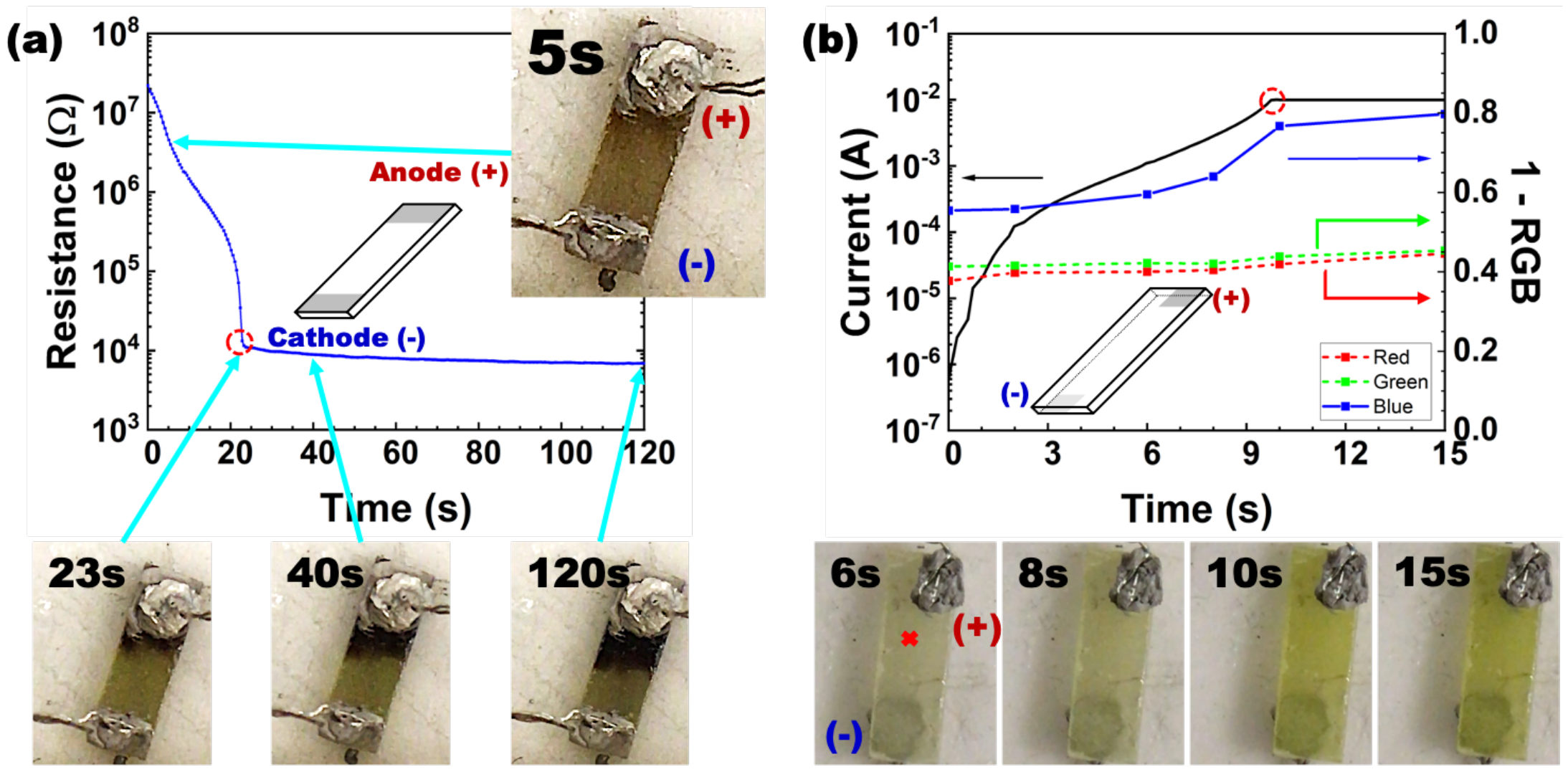

**Figure 3** (a) Resistance vs time plot of Fe-doped STO crystal degraded at 350°C and 200 V in air showing anode-initiated dark band developing much slower than resistance degradation. Inset: schematic top view of sample, electrodes shown as shaded areas. (b) Current (left scale) vs time plot of undoped STO crystal degraded at 350°C and 200 V in air. (Right scale) 1 minus average tristimulus value of each color in RGB at sample spot marked (red cross), yellow quantified by "1 − blue" given almost constant green and red values. Increasing yellow electroluminescence with current also evident in time-stamped photos. Inset: schematic top view of sample, half-width electrodes shown as shaded areas with cathode on bottom face and anode on top. Red circles in both plots mark start of compliance.

Compared to YSZ, the temperature dependence of $t^{\mathrm{f}}$ is weaker in STO. When $R(0)$ of Fe-doped STO increases from ~1 MΩ at 400°C to ~100 MΩ at 290°C, its (200 V) degradation time changes from a few seconds to a few minutes; in YSZ a similar $R(0)$ difference changes the degradation time from seconds to hours or days. Likewise, while $t^{\mathrm{f}}$ is just a few seconds at 400°C in both materials, at 290°C it increases by ~35× in STO compared to 500× in YSZ. Undoped STO degrades similarly, though typically faster (**Figure 3b**). Both crystals show considerable variance similar to what Waser *et al.*[2-3] found.[25-A5]

## Electrocoloration vs electroluminescence

Different regions in an *n-i-p* multi-junction appear as different color bands in **Figure 3**. With the reduction of residual $Fe^{4+}$, the cathode side gradually loses its initial light brown color once the test begins. Meanwhile, when $Fe^{4+}$ forms in the *p*-region, the anode side acquires a dark brown/black color, to be referred to as electrocoloration. After prolonged degradation, especially under a current compliance, $V_O^{\bullet\bullet}$ electromigration towards the cathode leaves the anode side "totally" $V_O^{\bullet\bullet}$ deprived, and such extreme oxidation manifests as a pitch-black color. With only 1/500$^{th}$ of the Fe concentration, undoped STO blackens much less (**Figure 3b**).

Just as Waser *et al.*[3] reported, the cathode side in **Figure 3a** brightens, turning bright yellow[12] during testing, which they attributed to $Fe^{3+}$. However, yellow color also appears in **Figure 3b** in undoped STO suggesting it is not from $Fe^{3+}$ but from the host, and the fact that even the anode side is yellow further rules out $Fe^{3+}$. (In doped STO, the yellowness of the anode side is presumably masked by too many $Fe^{4+}$.) Moreover, **Figure 3b** reveals that the yellowness is strongly correlated to *J*. (Color perception is known to scale with log(intensity), so the data indicate yellow luminescence is proportional to *J*.) Such *J*-luminescence correlation was seen in every sample we tested—doped and undoped ones, and in every phase—including when compliance control was entered[25-A6] and when *J* was suddenly quenched upon a metal-insulator transition. (See **Degraded STO undergoes a metal-insulator transition, Cooling in silicone oil: bubbles, flashes, luminescence and conductivity transition.**) Therefore, what was seen must be electroluminescence.

**DC electric degradation ends in a flash**

One of the most spectacular observations of DC degradation in 8YSZ was that it ended with a flash, starting as a bright, red or white spot at a point (e.g., a protrusion) right next to the anode. Some common requirements for the flash are (a) sample is under a constant voltage, (b) there is a point of localized and excessive Joule heating, which is possible if (c) the point is a conductivity minimum, and (d) the conductivity minimum spans a vanishingly thin/small spatial extent. These conditions will ensure a high enough current will arise to generate such concentrated Joule heating that thermal runaway must take place. Indeed, in 8YSZ, the last bit of the most resistive section is always right next to the anode, which is why flash occurs there.

In STO, DC degradation also ended in a flash if a current compliance was not set (thus meeting requirement (a)), and like in 8YSZ, a 10 mA or so compliance can control the flash to make it last; a smaller compliance will suppress it. However, flashes in STO do not start at or next to any electrode and their locations depend on the voltage. For example, at 350°C, the brightest part of the flash at 125 V is closer to the cathode in **Figure 4a**, but in **Figure 4b**, at 150 V, it is closer to the anode.

According to **Figure 1a-b**, the ionic "plug", which is the most resistive part (thus meeting requirement (c)) in a multi-junction, is not next to either electrode. (In conductivity bifurcation, the most resistive part should never be next to either electrode.) This agrees with our observations. More specifically, we expect the flash to occur at $x^{\mathrm{f}}$ when the *i*-region is being pinched off, because it satisfies requirement (d). However, Wang *et al.*[11] also found a higher voltage causes $\left[\mathrm{V_O^{\bullet\bullet}}\right]^{\mathrm{f}}$, hence $x^{\mathrm{f}}$, to increase, moving the ionic plug—hence the flash—toward cathode, which is opposite to what we observed. As already suggested in **Defect chemistry and conductivity** **bifurcation**, a severe oxygen loss at higher voltages can cause the *n*-region to grow and the ionic plug to move toward the anode. Such evolution, which can explain our results, is schematically illustrated in **Figure 4c-d.**

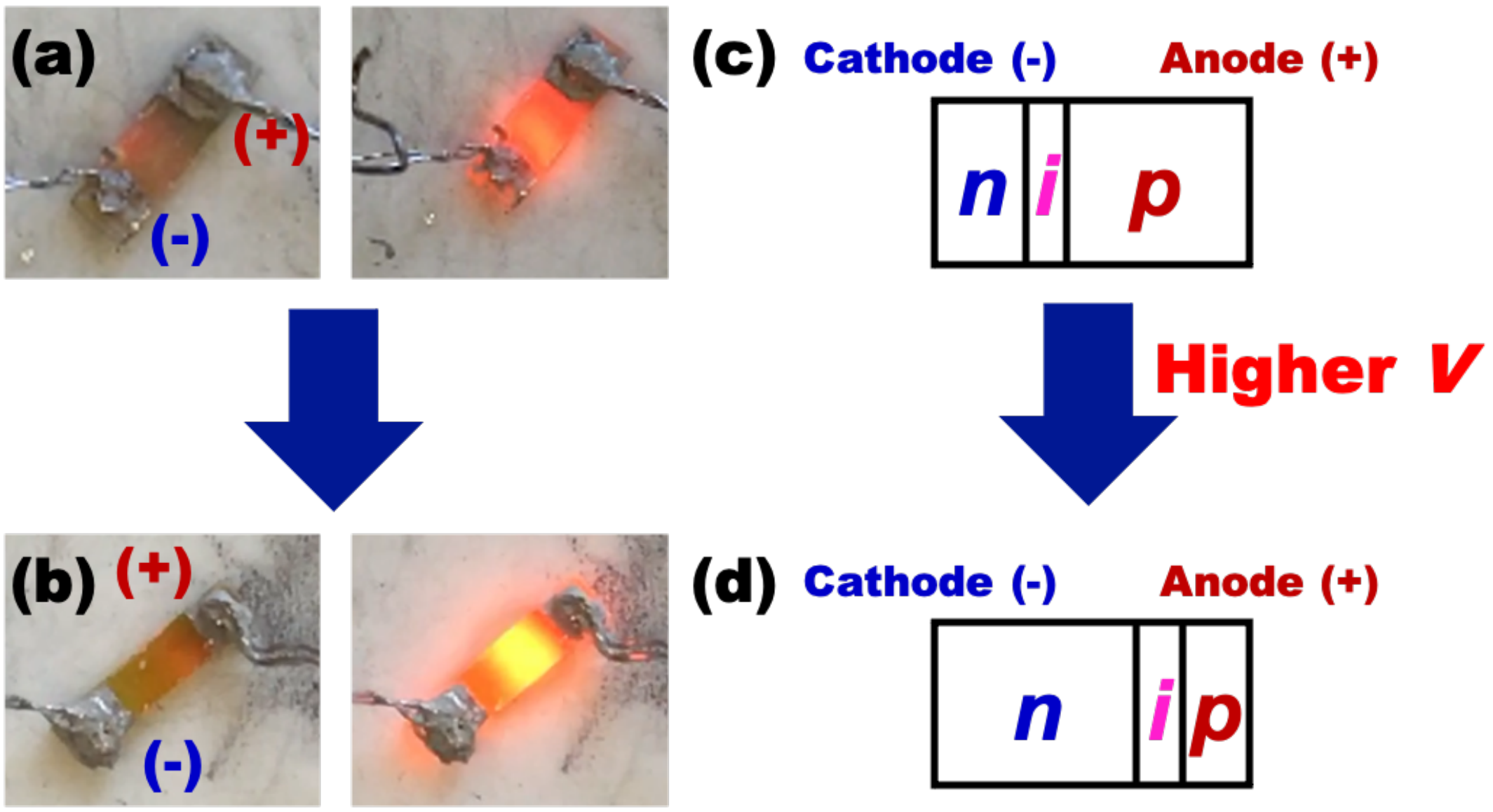

**Figure 4** Red spot in undoped STO near (a) cathode at 125 V and (b) anode at 150 V, both developing into a flash at 350°C in air. Schematic of ionic plug, closer to (c) cathode at lower voltage, and to (d) anode at higher voltage.

**Degraded STO undergoes a metal-insulator transition**

Severely DC-degraded 8YSZ are metallic but less degraded ones undergo an irreversible metal-insulator transition during cooling. The same happens to STO in **Figure 5a**, showing two degraded samples, held for 200 s after reaching the compliance, then cooled. Under 10 mA compliance, the resistance is flat within a factor of 2 during cooling, but under 5 mA, a sudden metal-insulator transition is signaled by a >100× resistance increase at ~87°C.

Because both samples were degraded at 200 V, which is a higher voltage than used in **Figure 4**, very likely their *n*-regions have nearly overtaken the entire sample sometime after reaching the compliance. This is especially true if oxygen loss has occurred, a distinct possibility at such voltage as suggested earlier (**DC electric degradation ends in a flash**) and supported by the decrease of voltage (from 200 V to <100 V) during the 200 s hold under the compliance before cooling began in **Figure 5a**. Before cooling, the *n*-region has a very high $\left[V_O^{\bullet\bullet}\right]$, hence $\left[e'\right] \approx 2\left[V_O^{\bullet\bullet}\right]$ there; since $V_O^{\bullet\bullet}$ is not very mobile and is easily quenched, we expect $\left[e'\right]$ to remain constant during cooling. Therefore, when the sample resistance is dominated by the *n*-region, the voltage should decrease with decreasing temperature to reflect the higher electron (and hole) mobility at lower temperatures[28] (see the 10 mA case at below 50°C), although it may initially increase (from 350°C to 50°C) most likely because of electron trapping at $V_O^{\bullet\bullet}$. The above considerations ruled out the *n*-region as the place of metal-insulator transition, leaving the *p*-region as the only possibility. This is plausible because at low (but fixed) $\left[V_O^{\bullet\bullet}\right]$, cooling below a ($\left[V_O^{\bullet\bullet}\right]$-dependent) temperature $T_{\text{critical}}$ as shown in **Figure 5b** can cause a rapid decrease of $\left[h^{\bullet}\right]$, thus triggering carrier localization. Presumably, the transition only happened in the 5 mA sample because it suffered less oxygen loss, thus had a thicker *p*-region, making hole resistance feature more prominently in the sample resistance. We will revisit this subject after repeating the tests in silicone oil (**Atmospheric oxygen affects electrocoloration and *R*(*t*), Cooling in silicone oil: bubbles, flashes, luminescence and conductivity transition**). Obviously, if an *i*-region, or a thick *p*-region in a *p-n* junction, remains, then the sample is an insulator. One example is shown in **Figure 5c** for a sample degraded to Stage II, then cooled.

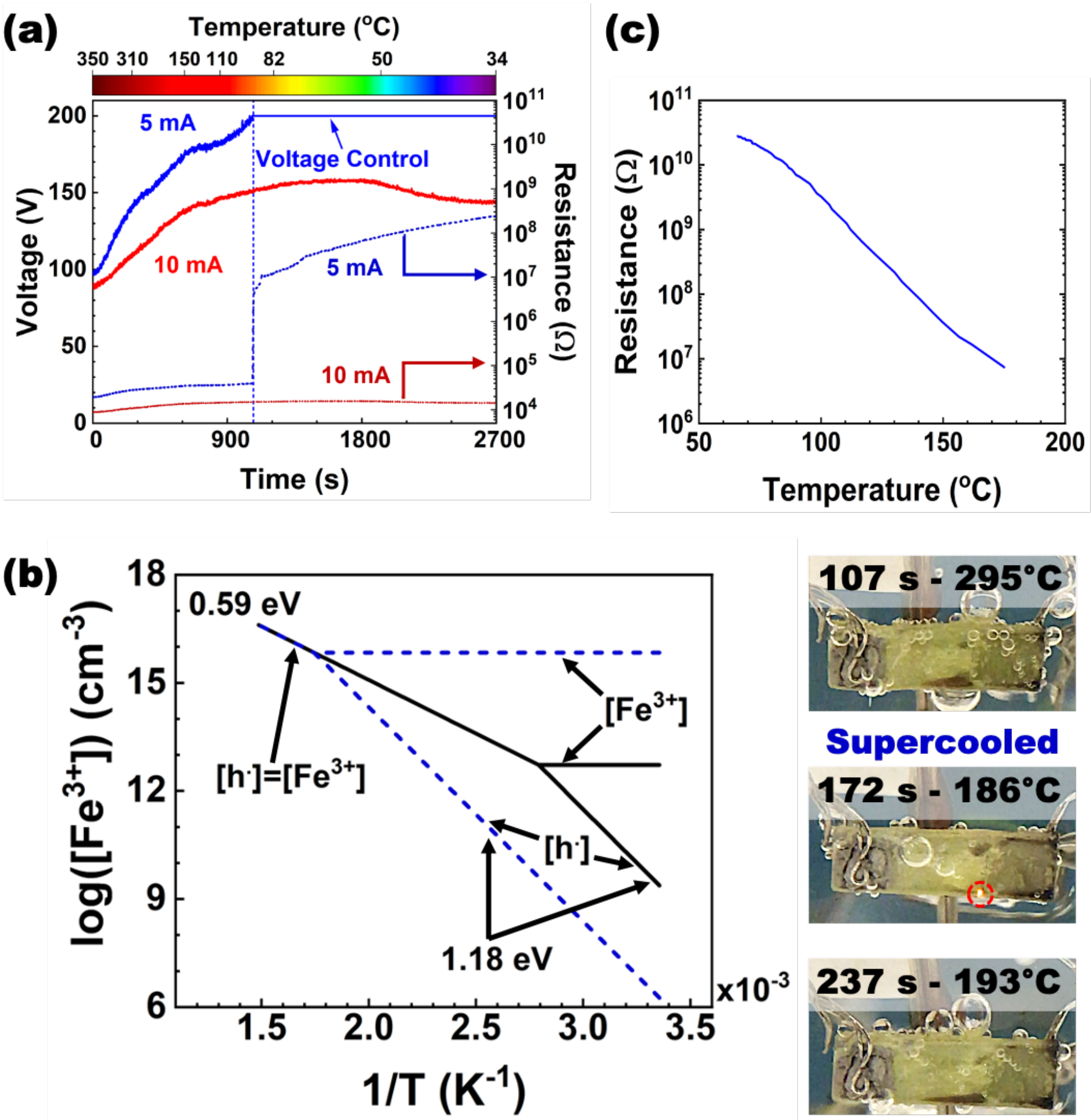

**Figure 5** (a) Voltage-time (solid curves) and resistance-time (dotted curves) plots of Fe-doped STO crystals degraded in air at 350°C and 200 V to enter 5 mA (blue) or 10 mA (red) current compliance, then after another 200 s (set as 0 s in plots) cooled to temperature shown in upper scale under same compliance until metal-insulator transition, upon which compliance is replaced by voltage (200 V) control. (b) At fixed $\left[V_O^{\bullet\bullet}\right]$ ($2.6\times10^{12}$ $cm^{-3}$ for black and $3.4\times10^{15}$ $cm^{-3}$ for blue) consistent with low $M_{V_O}$, temperature dependence of calculated $\left[h^{\bullet}\right]$ and [$Fe^{3+}$] changes abruptly at $T_{\text{critical}}$ in highly oxidized *p*-region in 0.05 wt.% Fe-doped STO. Below/above $T_{\text{critical}}$ is same as extrinsic/saturated regime in **Figure 1a**. Photos demonstrate redox effects of temperature-sensitive [$Fe^{3+}$]: time-sequenced snapshots taken from silicone-oil test (described in **Figure 12b** caption) showing blackened wedge, present at 295°C and 193°C, momentarily disappeared during interim flash to red hot state (red-circled) in supercooled sample. (c) Resistance-time plot of Fe-doped STO crystals degraded to prolonged Stage II in air at 200°C and 200 V, then cooled under 200 V.

**Resistance degradation faster than electrocoloration**

It was not uncommon to find one or more dark bands, often rather sharp and narrow, to span from cathode to anode in barely degraded (10-20% drop in resistance) 8YSZ crystals.[26] Dark bands in STO seemed to come much later. The sample in **Figure 3a** is black in less than 50% of the length even though it is well past the current compliance. Other samples just before reaching the compliance usually had their dark bands confined to within ~0.2 mm from the anode. Therefore, unlike in 8YSZ, resistance degradation in STO seems to proceed much faster than electrocoloration.

The observation in **Figure 3a** only painted a partial picture, however, and an examination of the back face, shown in **Figure 6**, revealed blackening started as soon as the test began. Yet its sight was blocked by the electrode in **Figure 3a** because it had not propagated much past anode's leading edge. The dark band in **Figure 6** was initiated relatively uniformly over most area beneath the anode, and it more or less grew as a plane-front parallel to the leading edge of the surface anode in agreement with **Figure 3a**. Such widespread darkening beneath the anode implies said electrode is surrounded by a uniformly resistive *p*-region, which makes it unlikely to have any anode-based field concentration to nucleate a localized dark band. Note that back-face examinations of 8YSZ also found its anode covered by a uniformly insulating region,[26] and in this respect the two materials behave similarly.

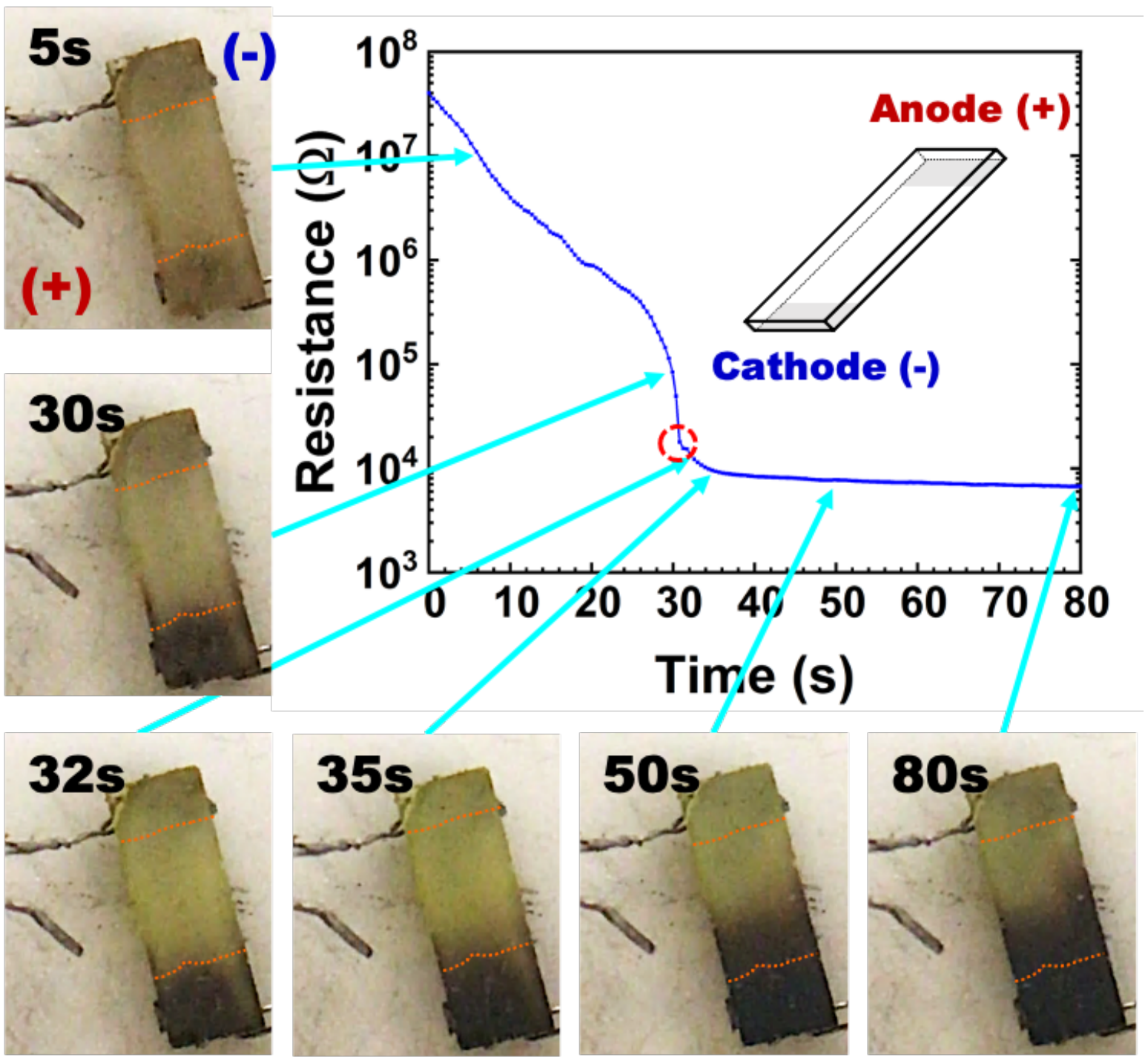

**Figure 6** Resistance vs time plot of Fe-doped STO crystal (viewed from back side) degraded at 350°C and 200 V in air showing dark band beneath anode. Darkening initiated long before reaching current compliance (red circle) along edges. Leading edge of surface anode marked as orange dotted line. Inset: schematic view of sample, electrodes being shaded areas. Red circle in plot marks start of compliance.

These observations suggest that, unlike in 8YSZ, one should not expect finger-like growth of fast-conducting paths in STO. This is despite their other similarities: Their cathode-side regions are highly conducting, even metallic; their anodes surrounded by an insulating region. On the other hand, the different impressions may simply reflect the difficulty of seeing fingers in STO, for any cathode-initiated finger is likely to be colorless, whereas cathode-initiated fingers in 8YSZ are black. To overcome the difficulty, we exploit other test environments next.

**Atmospheric oxygen affects electrocoloration and *R*(*t*)**

With no flash, Fe-doped STO samples held at a compliance current gradually became mostly black as shown in **Figure 7**, suggesting the *p*-region had spread to the entire sample. This would be a total surprise to the consensus theory, which envisions an *n*-region, not a *p*-region, to dominate in a completely degraded sample! Fortunately, repeating the test in silicone oil solved the puzzle: A similar $R(t)$ was obtained for a mostly clear sample in **Figure 8**, showing yellow luminescence.

As mentioned before, Stage II can be very long at lower temperatures, see **Figure 9a**. Typically, it happens after the sample has lost 99% of the initial resistance, and such sample, when cooled, shows a steep insulator-like resistance increase as in **Figure 5c** meaning it must have a substantial *p*-region. This is consistent with the remaining 500 kΩ-1 MΩ resistance of such samples, suggests a $10^{-5}$-$10^{-6}$ $\Omega^{-1}$ $cm^{-1}$ conductivity which is likely due to a (not yet fully oxidized) *p*-region. (Similarly, Wang *et al.* reported ~$10^{-6}$ $\Omega^{-1}$ $cm^{-1}$, at 210°C .)[11] When tested in silicone oil, such sample remains mostly clear except beneath the anode in **Figure 9b** (top and middle). HALT in air at 240-290°C also features an extended Stage II but often with a shallow resistance minimum, one shown in **Figure 9a**. Meanwhile, the sample is typically fully black after testing (**Figure 9b,** bottom). We believe the small gradual resistance increase past the resistance minimum is caused by reoxidation in the *n*-region, which increases its resistance. Usually, oxygen ingress is only possible at the cathode because the external circuit needs to supply 4 electrons to convert $O_2$ to 2 $O^{2-}$. But it is also possible on the surface of a conducting STO which can receive electron supply from the cathode. This explains why (a) there is no resistance increase in the silicone oil test, for it lacks oxygen supply, and (b) the *n*-region of air-tested STO becomes totally black, for atmospheric oxygen can enter this region thus annihilating $V_O^{\bullet\bullet}$ and restoring $Fe^{4+}$. Interestingly, samples tested in silicone oil in **Figure 9b** reveal two opposite developments in Stage II: the color of the darkened region becomes lighter while at the region's base next to the anode, it becomes darker. We will return to this point later (**Steady state possible in HALT?**).

In short, testing in silicone oil can avoid several reoxidation artifacts and, most importantly, extraneous darkening that masks all other features. Therefore, we sought other observations in such tests, some eventually revealing *invisible* short-circuiting needles.

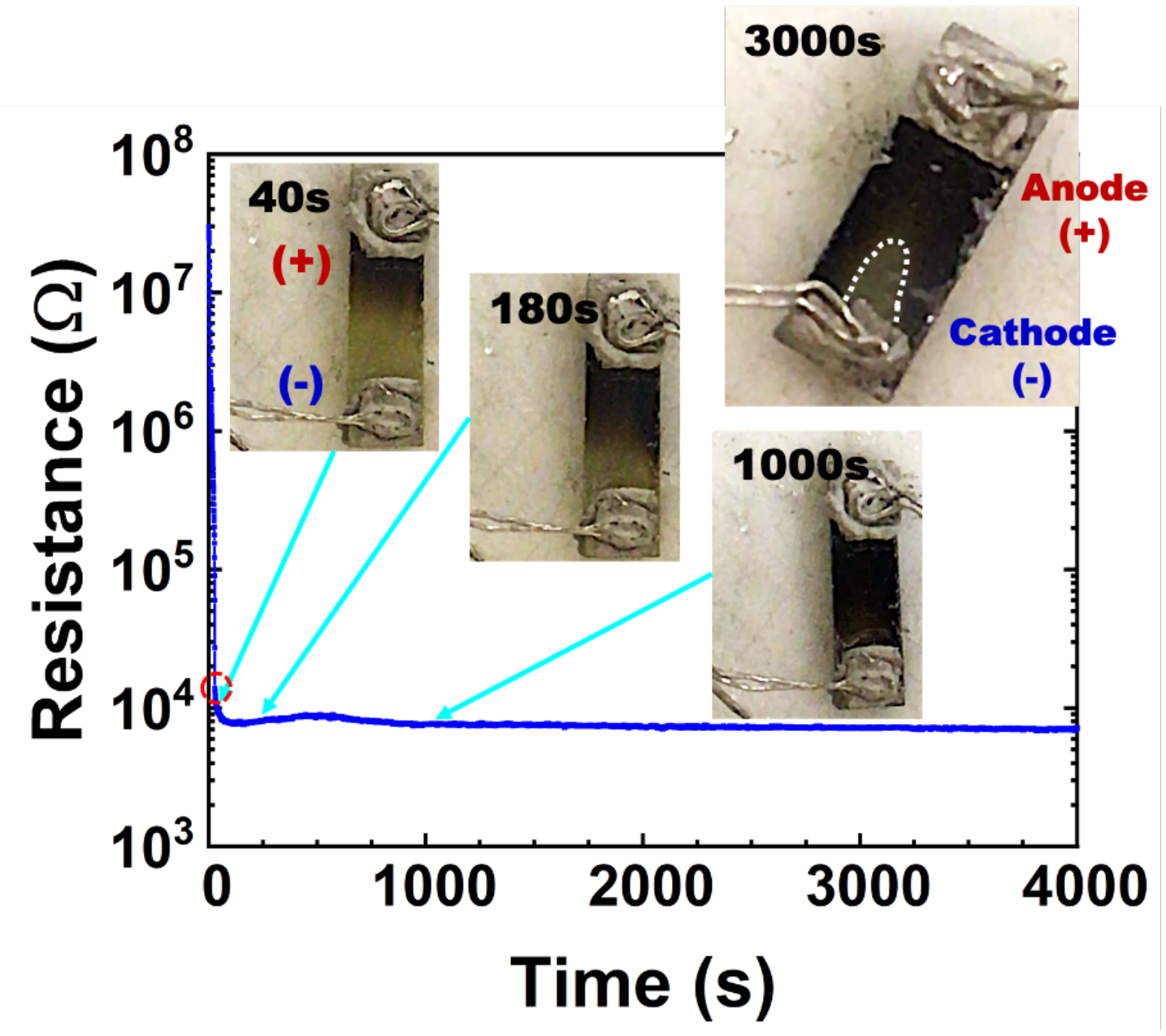

**Figure 7** Resistance vs time plot of Fe-doped STO crystal degraded in air at 350°C and 200 V under 10 mA compliance with dark band spreading to entire sample. Inset (upper right): Another Fe-doped crystal similarly degraded at 350°C for 3000 s showing full darkness except in outlined region. Red circle marks start of compliance.

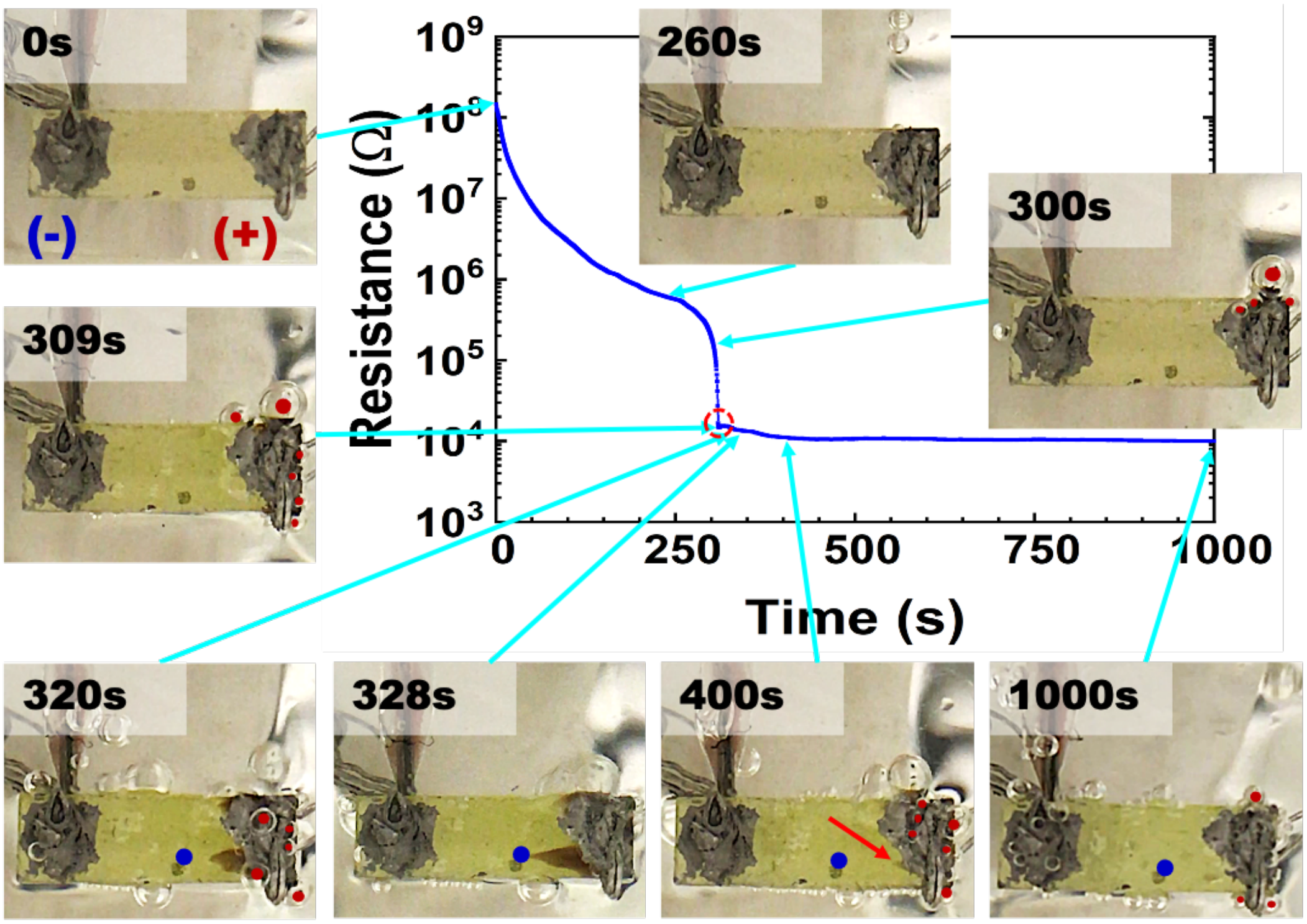

**Figure 8** Degradation curve for STO crystal in silicone oil at 290°C and 200 V with time-stamped photographs showing anode-emitted bubbles and creation of dark wedge (320 s), later nearly disappeared (400 s). Red circle on curve marks start of compliance, blue dots mark the (same) farthest spot reached by dark wedge, and red dots mark some anode-emitted bubbles.

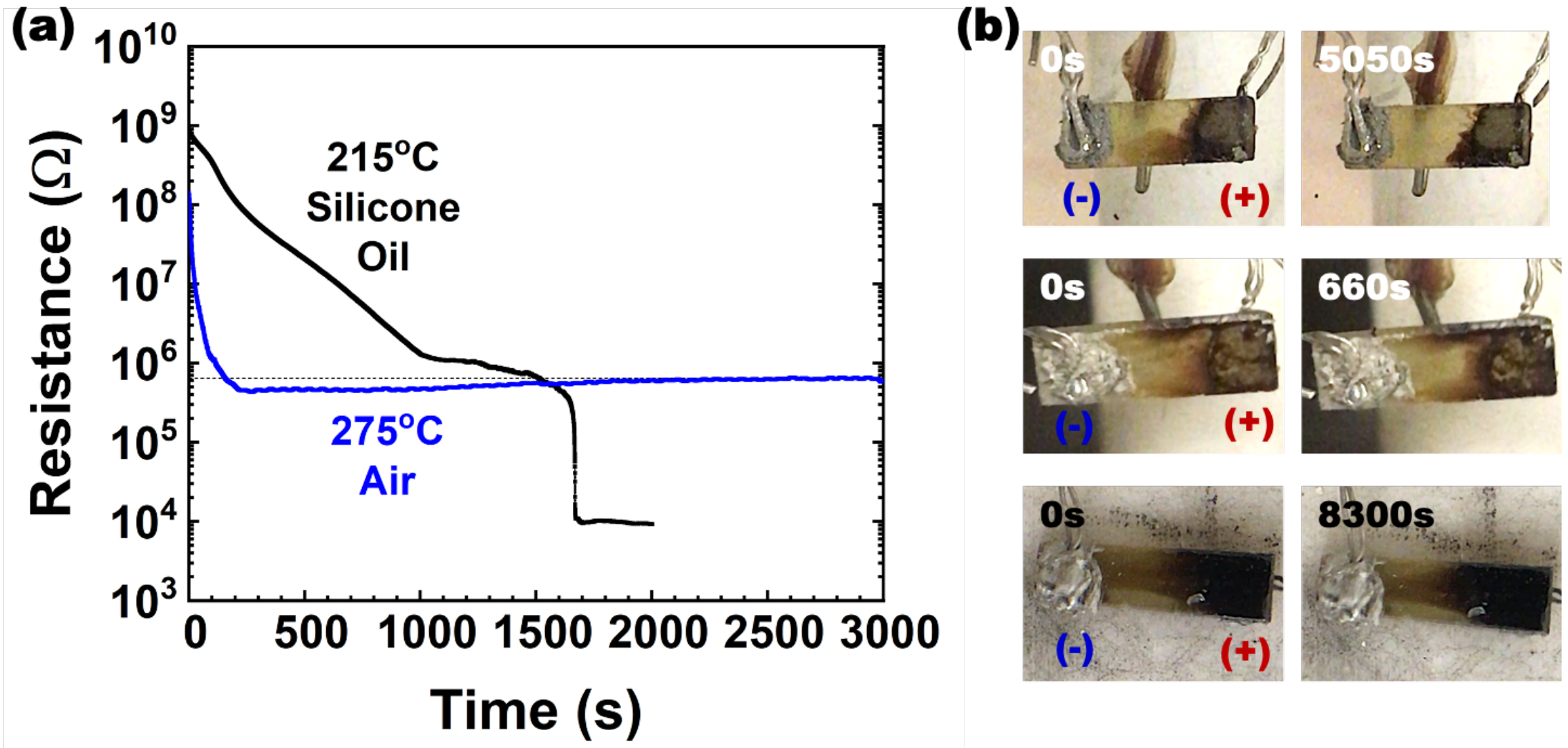

**Figure 9** (a) Resistance vs time plots of Fe-doped STO crystal with prolonged Stage II at 200 V, in silicone oil at 215°C and in air at 275°C. (b) Photos of Fe-doped STO crystal degraded at 200 V in silicone oil at (top) 205°C and (middle) at 215°C, and (bottom) in air at 215°C. Time stamps mark time in Stage II.

**Anode and "non-electrode" oxygen bubbles**

Starting at 300 s, oxygen bubbles marked by red dots in **Figure 8** begin to form at anode. This is right after $R(t)$ started a Stage III plunge, but well before the current (1 mA) had reached the set compliance (10 mA). Upon reaching the compliance, bubbling accelerates and continues thereafter. (Ignore the two off-sample bubbles at 260 s; they came from the bottom of the container where a hot spot caused silicone oil to bubble.) Since detection necessarily lags formation, and sample reduction with increased conductivity must follow bubble emission, we suggest oxygen bubbling is the cause of accelerating degradation in this stage: the plunge of the $R(t)$ curve, the onset of current compliance, the intensifying yellow electroluminescence accompanying the above (see the difference between 300 s and 309 s), and the further resistance drop afterwards. When kinetics is too slow and oxygen bubbling is postponed, Stage-III degradation is also postponed as seen in the low-temperature HALT in **Figure 9a**.

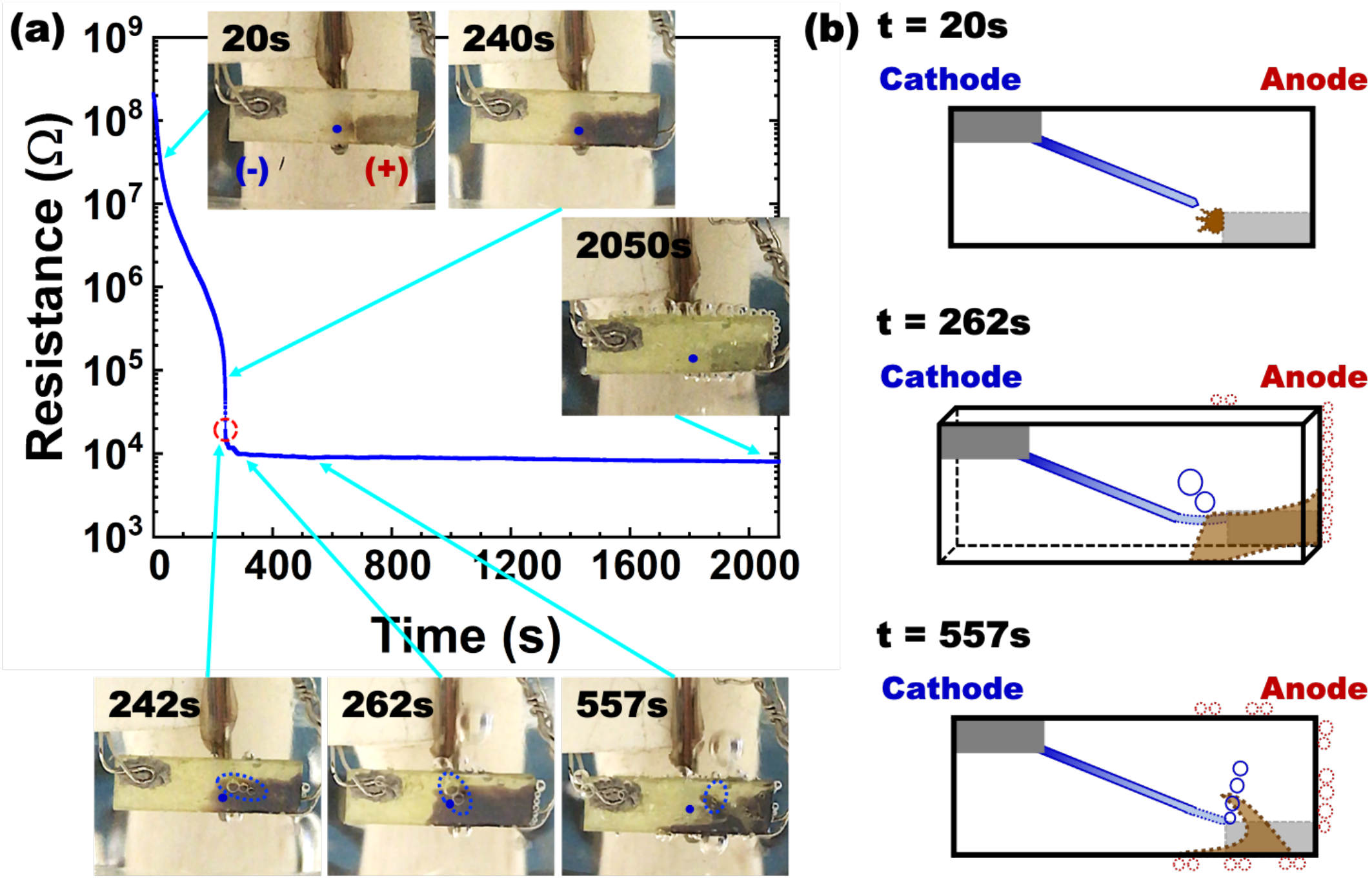

**Figure 10** (a) Resistance vs time plot of Fe-doped STO crystal degraded at 280°C and 200 V in silicone oil, with electrodes across body diagonal (half-width electrodes on opposite faces and corners) and sample viewed from top (cathode) face. Dark band started beneath (opposite face) anode from which numerous anode bubbles emerge at edges of bottom (anode) face. Additional non-electrode bubbles, encircled in dotted blue, were emitted from top face next to an invisible virtual-cathode needle, its likely tip location marked by blue dot. (b) Interpretive schematics for three photographs in (a) showing (top) needle-growth, (middle) dark-band progression and bubble formation of anode origin (dotted red circles) and non-electrode origin (blue circles), and (bottom) dark-band recession. Middle panel is 3D schematic, top and bottom panels are 2D projections. Dotted-broken lines or objects are outlines of objects beneath top face, vs solid ones for objects on top face.

**Figure 10a** shows a sample with two diagonally placed electrodes at two corners of opposite faces: cathode on the upper left of the top face, anode on the lower right of the bottom face. While it is impossible to see through Pt and monitor the anode surface, once anode bubbles migrate to the sample/electrode edges, they become visible as "edge bubbles" (unmarked in **Figure 10a**) starting at 242 s. Interestingly, "non-electrode bubbles" (encircled in dotted blue at 242 s, 262 s and 557 s) also form on the top face from a small region far away from any electrode, making us believe that they are emitted from the top STO surface. This is possible if we assume an *invisible* virtual electrode there: a virtual electrode initiated from the cathode and propagated along or near the top face to land its tip at the (same) blue dot (tentatively assigned, to be justified later) in **Figure 10a**. Such tip has a highly negative potential, which can strongly repel $O^{2-}$ thus building, against the top face, an $O^{2-}$ pileup—if bubble emission is initially inhibited—until it becomes strong enough to emit bubbles. (If there is some $Fe^{4+}$ in the vicinity, then the 4 electrons left by converting 2 $O^{2-}$ to 1 $O_2$ will go to them, thus no blackening is expected.) This picture is schematically depicted in **Figure 10b** and elaborated below by calling attention to other evidence in **Figures 8** and **10**.

### Virtual-cathode needles and their *E*-field lines

Envisioning a hypothetical virtual cathode as an *n*-type "needle" with converging *E*-field lines at its tip leads us to the following observations. (i) The field pushes away $O^{2-}$, which accumulates upstream and converts $Fe^{3+}$ to $Fe^{4+}$, causing blackness along field lines. (ii) If a high density of field lines intersect the sample surface, then they will cause an $O^{2-}$ pileup against the surface barrier and eventual emission of non-electrode bubbles. (iii) If the tip is within a characteristic length (either the width or the length) of surface anode, then its image charge will modify the field lines to make them fan out like a triangle, with its base falling on the front edge of the surface electrode and its tip pointing toward the needle. One such triangle, shaped as a wedge and blackened by $Fe^{4+}$, is seen at 328 s in **Figure 8**. (It grew from 309 s to 328 s, then shrank to almost disappearance at 400 s presumably because continual oxygen loss made the sample too severely reduced.) (iv) When both electrodes are on the top face, as in **Figure 8**, most field lines should lie on or near the top face, as does the needle. Some stray fields, though, will "dig below" to reach the bottom-face anode, which explains the blackness beneath the anode in **Figure 6**.

These field lines can also explain several other observations in **Figure 10**. (i) A blackened zone appearing at 20 s has a more diffuse appearance because the field lines there, presumably emerging from a needle tip on the top face and ending at the bottom-face anode, must traverse across the sample and suffer more optical scattering. (ii) The field lines in (i) being non-planar must shape like a cone with its base lying on the bottom electrode. This explains why, at 240 s, the black zone has covered the entire back side of the anode, similar to what is shown in **Figure 6**. (It is also a testament that the material around the anode must be insulating enough to support a diffuse electrical field. Otherwise, a short-circuit would have formed there.) (iii) Those field lines that intersect the top face are not far from the tip of the virtual cathode, also at or near the top face. So, they are rather intense and can push the $O^{2-}$ pileup at the top face to emit non-electrode-bubbles starting at 242 s. (iv) Oxygen outflow along the most intense field lines can cause reduction and remove $Fe^{4+}$ electrocoloration. It begins in front of the needle tip, clearing the blackness there. Meanwhile, the surrounding region that sees less intense field lines and less oxygen outflow is less reduced or not reduced at all, so it remains black. As a result, at 557 s the previous all-black region has a new appearance: It is like a set of V-shaped "jaws," "spitting" out non-electrode bubbles. (v) Finally, at 2050 s, reduction has become so widespread that no $Fe^{4+}$ remains, which eliminates all blackness except perhaps at one far section of the anode edge.

While the above was observed in silicone oil, air-tested samples most likely also emit oxygen bubbles in Stage III and continue bubbling when under compliance, during which there is further resistance drop that may be attributed to oxygen loss. Therefore, it is understandable why, in the late stage of compliance-controlled testing, they too develop a highly reduced region, such as the clear region outlined in dotted white in the inset of **Figure 7**. Presumably, this region is being pointed at by a virtual-cathode needle, which directs local reduction.

**Wavy black front in air-tested samples**

While tests in silicone oil have provided vivid redox evidence that implicates the presence of otherwise *invisible* needle-like virtual cathodes, there is also evidence of the same in air tested samples. Both samples in **Figure 11a-b** begin with an anode-side dark front that bears some black wedges or even black "needles", which we interpret as the outlines of field lines of one or more virtual-cathode needles. In this sense, they "mirror" the virtual-cathode needles in front of them. This is schematically depicted in **Figure 11c (top)**. Also depicted in **Figure 11c (bottom)** is a case when the density of virtual-cathode needles is so high that their "mirror" wedges begin to impinge, resembling a plane-front. We note that Havel *et al.*[12] saw a wavy black front in STO.

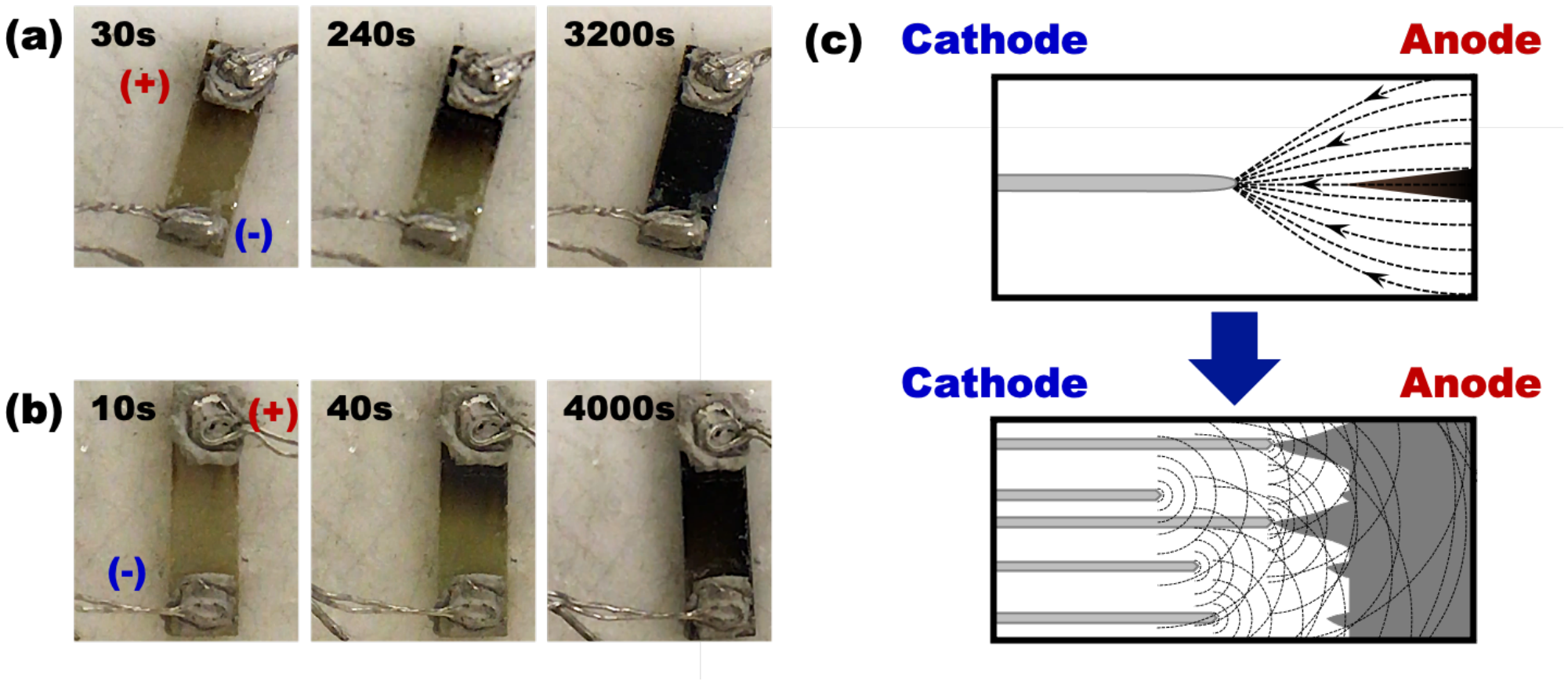

**Figure 11** Fe-doped STO crystals showing needle-growth turning to plane-front-growth after long-time testing at (a) 350°C and 125 V, and (b) 350°C and 200 V, both with 10 mA current compliance. (c) Top: Schematic field lines between tip of virtual-cathode needle at left and anode at right, covering approximately triangular area causing dark-

band growth from anode. Bottom: overlapping field lines from multiple virtual-cathode needles form plane-front-like black region with some pointed protrusions.

**Cooling in silicone oil: bubbles, flashes, luminescence and conductivity transition**

Two sets of Fe-doped STO were degraded to reach the 10 mA compliance, at 350°C, in silicone oil, and held them there for an additional 200 s before cooling under the same compliance in two ways.

(i) Slow cooling: With gradual addition of cold silicone oil to the bath, the cooling test confirmed a sudden resistance jump, by ~170×, at ~82°C in **Figure 12a**; anode bubbling also persisted down to 50°C despite a much smaller current. As expected, the sudden resistance increase quenched the yellow luminescence, visible before the transition at 2740 s and gone at 2790 s and 2920 s in **Figure 12a**, changing to pale white/colorless.

(ii) Supercooling: Using a glass tube to channel some cold silicone oil near the sample caused an almost immediate drop of the sample temperature, which later recovered partially when the warmer surrounding oil mixed in. Corresponding to the thermal transient was a resistance transient peaking in the interim: At its peak, a flash started at the most resistive spot in the sample, which was subject to a current compliance of 10 mA. As highlighted by red circles in **Figure 12b**, the first flash is a red-hot one appearing at 159 s (219°C), the second a white spot from 347 s to 352 s (101°C to 92°C). Note that both flashes occur at the same spot, right next to the anode. Shortly afterwards, at 88°C, a metal-insulator transition sets in most likely at the very same spot. Since the anode-edge is the most oxidized part of the sample and should have a higher $\sigma_h$ than further out, the fact that flashes occur there implies that the entire p-region has already shrunk to almost a point. Again, **Figure 12b** confirms (a) the resistance transition quenches the yellow luminescence (compare 352 s and 362 s), and (b) anode bubbling (see edge bubbles from 347 s through 362 s) continues to the lowest test temperature despite the insulator's miniscule current.

Lastly, in all but one photograph in **Figure 12b**, there is a blackened wedge located right in front of the flash point, e.g., at 347 s and 352 s. The wedge actually formed well before the compliance was hit. But curiously, while it existed at 107 s and 237 s as shown in **Figure 5b**, it disappeared in front of a red-hot flash at 159 s in **Figure 12b** and at 172 s in **Figure 5b**. The transitory disappearance/reappearance is caused by temperature-mediated $Fe^{3+}/Fe^{4+}$ redox in **Figure 5b**: Heating above $T_{\mathrm{critical}}$ during red-hot flashing causes reduction and wedge disappearance, and post-flashing cooling below $T_{\mathrm{critical}}$ causes reoxidation and wedge reappearance.

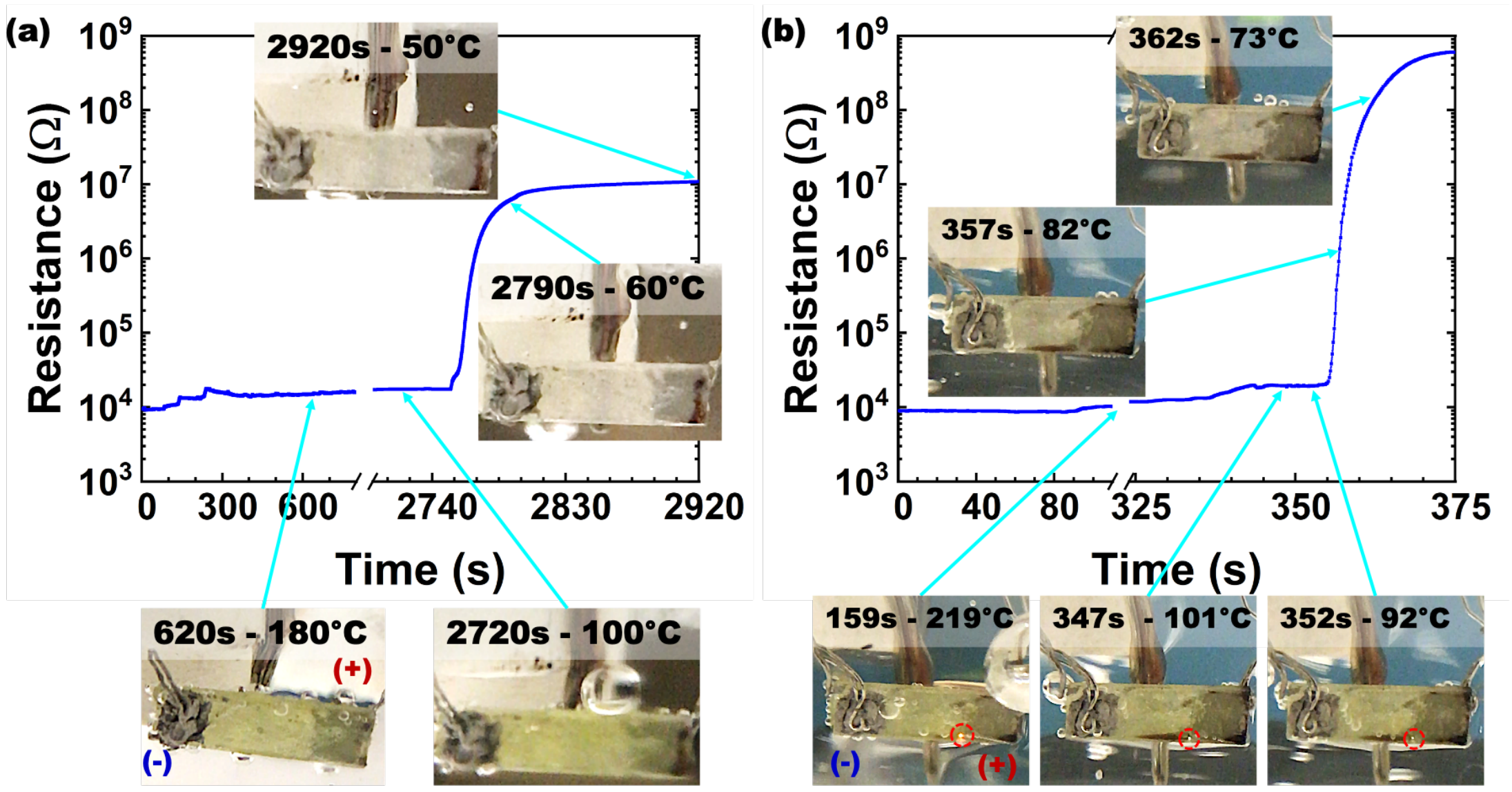

**Figure 12** Resistance vs time plot of two Fe-doped STO crystals, with full-width electrodes on opposite faces and in silicone oil, degraded at 350°C and 200 V until reaching 10 mA compliance, then remaining under same compliance until metal-insulator transition, upon which changing to voltage (200 V) control. Time 0 is when cooling begins (200 s after compliance). (a) In monotonic slow cooling, anode bubbling persists to 50°C, well below ~82°C of metal-insulator transition, which quenches yellow luminescence. (b) Faster cooling with two interim supercooling episodes, which under current compliance trigger flashes at red circles in time-temperature-stamped photos. Metal-insulator transition at ~88°C quenched yellow luminescence, but anode bubbling persisted to 73°C. Additional photographs for (b) illustrating redox effects in **Figure 5b**.

## DISCUSSION

### Needle-growth means strong *E* dependence

Our experimental evidence for needle-growth challenges the notion of plane-front growth in **Figure 1b**. Needle-growth mechanism is well accepted for dielectric breakdown of thin-film ceramics. In bulk-crystal HALT, we observed needles in both early and advanced stages of degradation (their $R$ varying by $10^3$-$10^4\times$), so it is also relevant. In retrospect, needle-growth should have been expected because in the Laplace equation of electrostatic potential, a plane-front solution is fundamentally unstable. Physically, any protrusion with an arbitrarily small local radius of curvature ($a$) on the plane front will consume almost all the potential drop (or driving force, in general), most of it within a radial distance $r \sim a$ from the tip of the perturbation. Since a sharp potential gradient equals a sharp field, it follows that the sharpest protrusion will grow the fastest, eventually into a needle with a tip radius $a$. Finger/needle-growth instability is well known in solidification[29] and diffusion-limited aggregation[30]; in electroreduction and DC degradation, it was also evidenced in 8YSZ,[26,31] and now in STO.

Unlike plane-front growth, needle-growth is not a one-dimensional (1D) problem. Nevertheless, the underlying physics—electromigration and electroreduction—is the same, though it must be solved around the tip of a needle using either spherical coordinates in 3D—for a tip buried inside a sample, or cylindrical coordinates in 2D—for a near-surface tip between two same-surface electrodes. Once again, an *n-i-p* multi-junction extended to the anode is expected, but it is one initiated not from the cathode but from the tip of the needle, and its $\Delta\phi$ is not expended over entire $L$ but mostly within $r \sim a$. This means a much more compressed *n*-region spanning $\sim a$ in size, which supports a much stronger field/flux, and it is surrounded by a large *i/p*-region spreading out like a fan (in 2D) or a conical cone (in 3D) while carrying only a relatively weak field/flux. As illustrated by the frames of 309, 320 and 328 s in **Figure 8**, it takes time to muster electromigration to develop the above multi-junction, and at 328 s the black triangle has reached its largest size. If we take it as the quasi-steady state, then the needle tip being a distance $a$ away must be right next to the tip of the triangle—for $a$ is very small. This allows us to mark the tip by a blue dot in **Figure 8**. (The triangle later retreats because of the overall reduction of the sample.)

Like the *n*-region in our plane-front solution, the needle being a virtual cathode is like a short circuit, so its tip potential is $\phi_{\text{tip}} = \phi_{\text{C}}$. The tip field, about $(\phi_{\text{C}} - \phi_{\text{A}})/a$, is thus much stronger than the nominal field $(\phi_{\text{C}} - \phi_{\text{A}})/L$ that drives the plane-front growth. However, the fact that we did observe "stuck" needles in **Figures 8** and **10** (the one in **Figure 8** stuck while still far away from the anode) suggests that $\phi_{\text{tip}}$ must have fallen after some growth, fallen so much that the remaining field, $(\phi_{\text{tip}} - \phi_{\text{A}})/a$, is too small to propel further growth. To understand why, let us assume that the initial growth occurs not by electromigration but by dielectric breakdown, via ionization and/or electron avalanche, which is the nanoscopic mechanism seen in very thin films. Unlike in thin films, though, athermal breakdown in a bulk sample will stop because the broken-down material left in its wake (the rear part of a needle) is highly defective and resistive. This makes a long, grown needle so resistive that its tip potential $\phi_{\text{tip}}$ must fall much below $\phi_{\text{C}}$, which explains why we saw long needle get stuck. This diffusionless mechanism can explain the extremely rapid initial needle-growth and the equally rapid initial $R(t)$ drop, falling 3-4 orders of magnitude in 30 s in mm-sized samples. Note that needle-growth instability pertaining to the Laplace equation is applicable to athermal breakdown—lightning in sky being one example.

The mechanism can also explain the strong *E* dependence of $t^{\text{f}}$. Athermal breakdown requires a threshold field; in bulk samples, it can only be reached when the relatively small applied *E*, $(\phi_{\text{C}} - \phi_{\text{A}})/L$, is hugely amplified by a field concentrator. Since the statistical availability of concentrators decreases with their potency (e.g., measured by their sharpness, $1/a$), a higher nominal *E* will find more concentrators that are potent enough to trigger breakdown. Therefore, an overall strong nominal-$E$ dependence is expected for $t^{\mathrm{f}}$, which explains the HALT data.

Polycrystals of STO are known to have a stronger $E$-dependence than single crystals.[2-3] In a polycrystal, needle-growth needs to restart every time it reaches a grain boundary. Thus, the search for the most potent field concentrators must be repeated many times, which accentuates the strong $E$-dependence. Grain boundaries are also known to be sites of defects, and grain size variation will further the disparity of defect distribution. Moreover, grain boundaries as kinetic bottlenecks are sites of large overpotential, hence stress concentrations.[4,32] These factors all conspire to make defect statistics more important in polycrystals than in single crystals, hence a stronger $E$-dependence. In this respect, porosity has the same effect of providing more defects, and it is well known that dielectric strength decreases with porosity.

Lastly, the large variation of $t^{\mathrm{f}}$ may also be related to needle-like growth. Recently, it was reported that dislocations can cause an increase in $\sigma_e$ under a reducing atmosphere at 650°C,[33] which could foretell the growth of surface-deformation-related needles. Indeed, darkened wedges along sample edges (in **Figure 5b**, **6**, **9b**, **11b** and **12b**) where surface deformation is most likely could mirror edge needles. At about the same temperature, a $PO_2 > 10^{-5}$ atm was also found to cause a decrease in $\sigma_{\mathrm{h}}$ and $V_O^{\bullet\bullet}$ mobility.[33] This could explain why $t^{\mathrm{f}}$ in air, with reoxidation during a prolonged Stage II at lower temperatures, is often much longer than that in silicone oil.

## Steady state possible in HALT?

A steady state must be nonchanging, including stoichiometry. Since oxygen is usually blocked at the cathode, no oxygen should exit the anode either, meaning no oxygen flow at all at the steady state. This demands a constant electrochemical potential of $V_O^{\bullet\bullet}$ and disallows any *i*-region, leaving *p*-*n* junction as the only option. Meanwhile, the sum of electron and hole currents must be constant, meaning the entire electric potential must be borne by the *p*-region because the *n*-region is much more conducting. From the above, we will argue below that, despite their claims, all the previous numerical solutions had not reached the steady state.[4,11-12,14-15]

To see why, we reproduce in **Figure 13** a "steady state" $\left[V_O^{\bullet\bullet}\right]$ profile from Fig. 6a of Wang *et al.*[11] The calculation was for 210°C and 0.8 kV/cm over 0.3 cm, giving a total voltage of 240 V. Dividing the *p*-region into two equal parts, from 0 to $x^f/2$ and from $x^f/2$ to $x^f$, and applying the condition of constant electrochemical potential for divalent $V_O^{\bullet\bullet}$, we obtain a potential drop of 0.129 V for ($x^f/2, x^f$), leaving almost the entire 240 V to ($0, x^f/2$). This is impossible because: (i) $\left[h^\bullet\right]$ (also shown in **Figure 13** though not to scale) and $\sigma_h$ are higher in ($0, x^f/2$) than in ($x^f/2, x^f$), and (ii) outside the *p*-*n* junction there is no high field (their Fig. 6e) in Wang *et al.*[11] A similar but lesser problem exists in Baiatu *et al.*'s calculations,[4] which used 0.01 kV/cm over 0.1 cm for a total voltage of 1 V, although the calculations were purported to represent their HALT experiments that mostly used 1 kV/cm or higher. The above argument also applies to single crystals and polycrystals because, at the steady state, the absence of oxygen current makes the afore-mentioned grain-boundary bottlenecks moot.

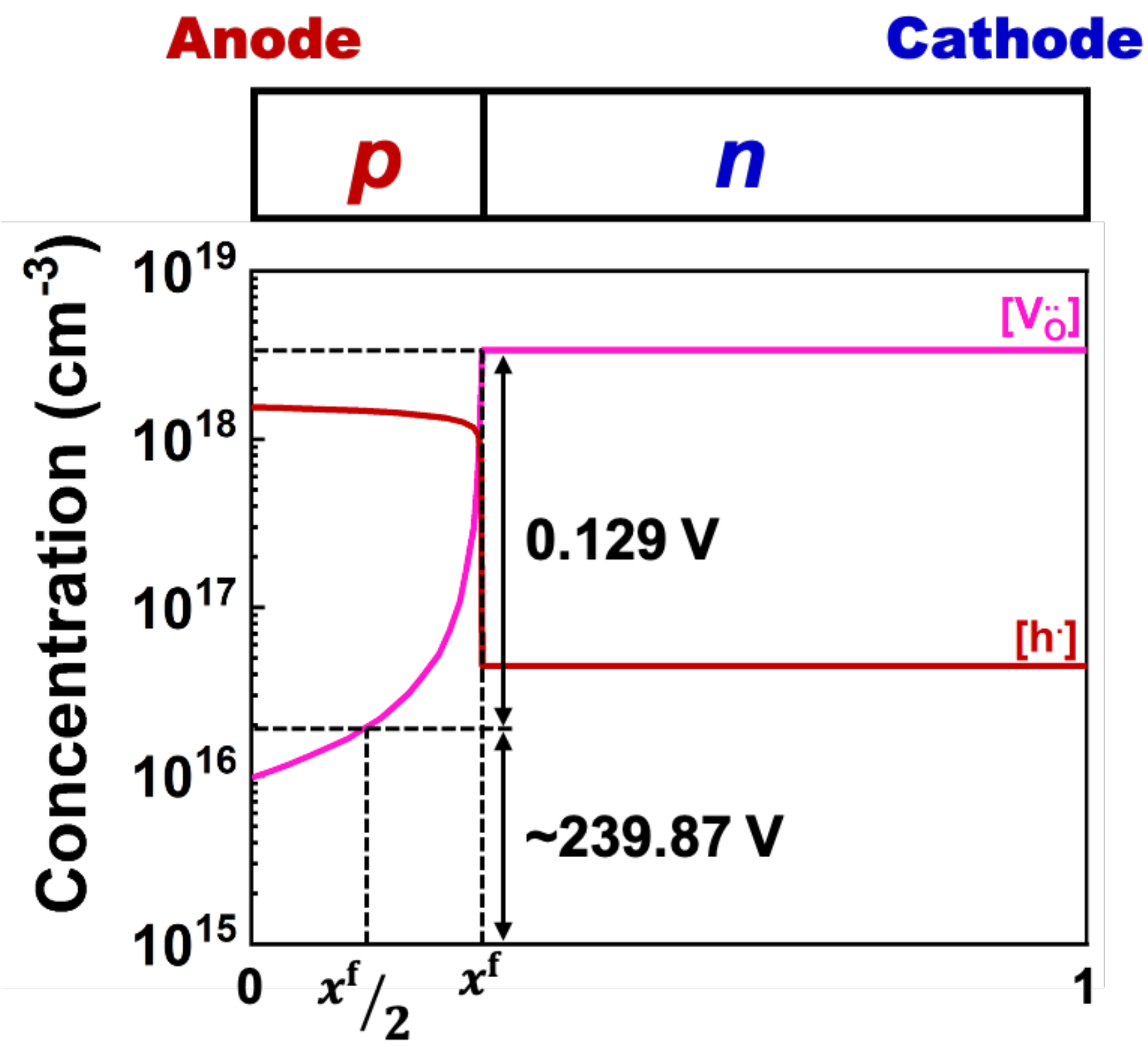

**Figure 13** Schematic *p-n* junction (top) with corresponding $\left[V_O^{\bullet\bullet}\right]$ (from Fig. 6a of Ref. 11) and $\left[h^{\bullet}\right]$ (not to scale) profiles. Calculated from $(kT/2e)\ln\left(\left[V_O^{\bullet\bullet}\right]^f/\left[V_O^{\bullet\bullet}\right]\right)$ where $\left[V_O^{\bullet\bullet}\right]^f$ is *n*-side-$\left[V_O^{\bullet\bullet}\right]$ at *p-n* junction at $x^f$, potential drop over $(x^f/2, x^f)$ equals 0.129 V, leaving remaining ~239.97 V to $(0, x^f/2)$.

Experimentally, a "steady-state" with a degraded but relatively stable resistance is often reported at lower HALT temperatures, two shown in **Figure 9**. (Wang *et al.*'s Fig. 5b reporting an effective conductivity of $10^{-6}$ $\Omega^{-1}$cm$^{-1}$ at 210°C is consistent with **Figure 9**.[11]) At about 210°C, the ratio of the final conductivity to initial conductivity in these tests is about 100. According to **Figure 1a-b**, this means a *p-n* junction has formed and $\sigma_h$ has taken control over $\sigma_{V_O}$. However, the *p*-region is not yet oxidized enough to reach $\sigma_h^{saturated}$, which would have given a conductivity ratio of $\sigma_h^{saturated}/\sigma_{V_O} \sim 2500$.

Therefore, we believe they are not the *true* steady state yet, and there must be a $V_O^{\bullet\bullet}$ flow from the *p*-region to the *n*-region, which will (a) oxidize the next-to-anode region further and (b) grow the *n*-region if $\left[V_O^{\bullet\bullet}\right]^f$ is fixed, both indeed seen in **Figure 9b**. In so doing, $E$ and $\sigma_h$ in the *p*-region will increase, and the sample resistance will decrease to approach the true steady-state value, about 1/2500 of the initial resistance. This is a slow process, however. As can be seen from Eq. (1), $d\ln\left[V_O^{\bullet\bullet}\right]/dt \approx -M_{V_O} z_{V_O} e\Delta\phi/L^2$ or $d\ln\left[V_O^{\bullet\bullet}\right]/d\left(t/t^f\right) \sim 1$, meaning each one-decade decrement in $\left[V_O^{\bullet\bullet}\right]$ will take about a time $t^f \ln(10)$, and there are still many decades to go according to the above analysis of potential drop. Therefore, the march to the true steady state may take a very long time in low-temperature HALT and its numerical simulation.

In the region of saturated $\left[h^\bullet\right]$ and $\sigma_h$, it is easy to show that oxygen potential follows $kT\ln\left(PO_2^{high}/PO_2^{low}\right) = 4e\Delta\phi$, which for $\Delta\phi = 240$ V comes to a $PO_2$ ratio of $10^{10017}$! This means an enormous $PO_2$ at the anode, which most likely will be forced to emit oxygen bubbles, as we observed in higher temperature HALT. (If a virtual-cathode needle has its tip pointing close to the STO surface, then a similar pressure will result at the surface and force non-electrode oxygen bubbles to emit there as well, which we also observed at higher temperatures.) Therefore, even with fast kinetics, the *true* steady state may still be elusive because the process of extreme oxidation of the near-anode region is likely interrupted by anode bubbling, which will supplant near-anode oxidation by general reduction. We believe this usually happens when the resistance ratio reaches about $\sigma_h^{saturated}/\sigma_{V_O}$. In our high-temperature HALT, such ratio is reached at the start of Stage III, which is when initial bubbling was witnessed. (Other than bubbling, anode-metal oxidation is possible albeit at the cost of reduced current density; e.g., Pt can form $PtO_2$ which is a good insulator capable of supporting a large voltage.)

The above considerations also apply to thin films. Because *kT* is usually less than 0.05 eV at any reasonable HALT temperature, even a 1 V bias is more than what a *p-n* junction in a steady state can bear. Allowing an additional electrode Nernst overpotential, typically ~1 V, we expect a bias of 3 V or more will yield the same outcome of anode bubbling and/or anode metal oxidation in the final stage of HALT in MLCC and thin films (like ReRAMs).

Finally, it is instructive to contemplate what the steady state solution should look like if the anode is infinitely strong and inert. Let us assume (a) $\left[V_O^{\bullet\bullet}\right]^f$ in the *n*-region is insensitive to $\Delta\phi$ as in our analytic solution, (b) the *p-n* junction potential $\Delta\phi^*$ is insensitive to $\Delta\phi$, (c) *n*-region's $\sigma_e$ is so high that there is hardly any potential drop there, (d) $\left[h^\bullet\right]$ and $\sigma_h$ right next to the anode—though not elsewhere in the *p*-region—have saturated, (e) a correct steady-state $\left[V_O^{\bullet\bullet}\right]$ profile albeit for a rather small $\Delta\phi$ is already available. Under these assumptions, the steady-state *p*-region $\left[V_O^{\bullet\bullet}\right]$ profile (called $\left[V_O^{\bullet\bullet}\right]^{new}$ below) for an arbitrarily large potential drop, $\lambda\left(\Delta\phi-\Delta\phi^*\right)+\Delta\phi^*$, where $\lambda$ is a constant, can be generated from the $\left[V_O^{\bullet\bullet}\right]$ profile in (e) by a transformation: $\left[V_O^{\bullet\bullet}\right]^{new}=\left[V_O^{\bullet\bullet}\right]^0\left(\left[V_O^{\bullet\bullet}\right]/\left[V_O^{\bullet\bullet}\right]^0\right)^\lambda$. Here, the notation in **Figure 1c** is used: $\left[V_O^{\bullet\bullet}\right]^0$ is the $\left[V_O^{\bullet\bullet}\right]$ at the *p*-edge of the *p-n* junction. Elsewhere in the *n*-region, $\left[V_O^{\bullet\bullet}\right]^{new}=\left[V_O^{\bullet\bullet}\right]^f$ according to (a).

The validity of the new steady-state solution is verified in four steps. (i) The electrochemical potential for $V_O^{\bullet\bullet}$ with $[V_O^{\bullet\bullet}]^{new}$ in the *p*-region is constant under the new potential. (ii) Since $[h^{\bullet}]$ and $\sigma_h$ right next to the anode have already saturated, they remain unchanged with or without the transformation, except for a hole current $\lambda$ times higher after transformation. (iii) Since $[e']$ and $\sigma_e$ in the *n*-region have already reached very high values requiring a negligible field to drive any arbitrarily large electron current, the same values suffice with or without the transformation, except for an electron current $\lambda$ times higher after transformation. (iv) Since $[V_O^{\bullet\bullet}]^{f}$ in the *n*-region is insensitive to field, the same value suffices with or without the transformation, thus still maintaining charge neutrality achieved by $[V_O^{\bullet\bullet}]^{f}$ and $[e']$ there.

To visualize $[V_O^{\bullet\bullet}]^{new}$, start by "anchoring" $[V_O^{\bullet\bullet}]^{new}$ at the same $[V_O^{\bullet\bullet}]^{0}$ at the *p*-edge of the *p-n* junction where the potential is $\phi_C + \Delta\phi^* = \phi_A - (\Delta\phi - \Delta\phi^*)$, then on the ln-plot give $\ln[V_O^{\bullet\bullet}]$ at other locations in the *p*-region a downward "affine-shift" of $(\lambda - 1)$ times $\left(\ln[V_O^{\bullet\bullet}] - \ln[V_O^{\bullet\bullet}]^{0}\right)$. For large $\lambda$, it brings to $\ln[V_O^{\bullet\bullet}]^{new}$ a value hundreds to thousands lower than $\ln[V_O^{\bullet\bullet}]$, which is needed to support a voltage of >10 V or >100 V. Note that other than the much smaller value of $[V_O^{\bullet\bullet}]^{new}$ (by many hundred to thousand orders of magnitude), the shapes of $\ln[V_O^{\bullet\bullet}]$ and $\ln[V_O^{\bullet\bullet}]^{new}$ are similar and not unlike that in **Figure 13**. This is because, given a constant electrochemical potential of $V_O^{\bullet\bullet}$ and to maintain a constant hole current in the *p*-region, $d\ln[V_O^{\bullet\bullet}]/dx \propto d\phi/dx \propto E \propto 1/[h^{\bullet}]$ must hold. The slope of $\ln[V_O^{\bullet\bullet}]$ in **Figure 13** meets the above expectation: (i) It approaches a constant right next to the anode corresponding to a constant $E$, just as it is supposed to for a saturated $[h^{\bullet}]$, and (ii) it becomes increasingly steeper as the *p-n* junction is approached corresponding to an increasing *E*, just as it is supposed to for a rapidly decreasing $\left[h^{\bullet}\right]$ there (not drawn to scale in the figure).

Incidentally, the steady state $PO_2$ distribution as well as $J^{ss}$ can be directly solved using the steady-state transport equation given in Ref. 32 and 34. Also note that the $J^{ss}$ obtained in the previous numerical solutions[4,11] is too low to account for our observations and most of Waser *et al.*'s.[2-3]

**Flash sintering**

In flash sintering, a porous powder compact is held under a constant and usually quite large voltage in an environment whose temperature is raised constantly. At a certain point the current density shoots up and the sample flashes—not unlike what is shown in **Figure 4**, allowing most porosity to sinter shut very rapidly. Before the flash, there is ample open porosity around most powder particles, making their oxygen-exchange with the atmosphere entirely possible. Therefore, $\left[V_O^{\bullet\bullet}\right]$ in such sample should be largely constant. This makes a porous STO similar to 8YSZ, even though STO is a model semiconductor whereas 8YSZ is endowed with an extremely large and stable $\left[V_O^{\bullet\bullet}\right]$ population. For the above reason, we do not expect any *p-i-n* or *p-n* junctions to form in porous STO before it flashes, that is, flash sintering of titanates can only be triggered by thermal runaway.[20,23-24] However, once flash densification is initiated and especially after open porosity is removed, ready oxygen exchange is no longer possible. From then on, all the (rather fast) phenomena described above for DC resistance degradation, especially the rapid current increase and anode oxygen egress upon the pinch-off of the *i*-region, are expected in a flash-sintered STO sample.

## CONCLUSIONS

(1) Definitive visual evidence has revealed the existence of virtual-cathode needles in all stages of DC resistance degradation of intentionally/unintentionally-Fe-doped $SrTiO_3$. Fast, flaw-statistics-dependent needle nucleation and growth can explain why, in highly accelerated tests, sample lifetimes are strongly dependent on the nominal electric field, a common observation that runs counter to the findings of previous numerical simulations.

(2) The total conductivity $\sigma_{total}$ of $SrTiO_3$ in **Figure 1a** is boat-shaped and the initial state is at the bottom of flat $\sigma_{total}$. Therefore, any $\left[V_O^{\bullet\bullet}\right]$-mediated carrier segregation will bring about conductivity bifurcation endowing more overall conductivity to the sample. Bifurcation is more severe at lower temperature and under a higher field, meaning more severe resistance degradation.

(3) Commonly reported end states of highly accelerated lifetime tests featuring a degraded but relatively stable resistance are not the steady state. The resistance will continue to degrade after an incubation time, by oxygen-vacancy outflow from the *p*-region to the *n*-region, causing the highly conductive *n*-region to grow and the near-anode region to further oxidize thus creating more holes.

(4) At higher temperatures, faster migration in accelerated lifetime tests leads to faster resistance degradation and near-anode oxidation. The process is likely truncated and reversed by the release of oxygen bubbles, hence general reduction, thus eliminating the *p*-region and leading to final resistance breakdown in $SrTiO_3$.

(5) Fully degraded $SrTiO_3$ with various degrees of oxygen loss is either an *n*-type metal or an *n*-type metal with a *p*-type residue next to the anode, the latter suffering a metal-insulator transition during cooling.

(6) When at least 3 V is applied to $SrTiO_3$, the above findings hold for single crystals and polycrystals, for bulk and thin ceramics, and for thin films. The findings also see analogy in DC degraded $Y_{0.08}Zr_{0.92}O_{1.96}$ despite entirely different defect chemistry and conduction mechanism.

(7) Yellow electroluminescence, its intensity positively correlated to the current density, is observed in both Fe-doped and undoped $SrTiO_3$. It serves as a visual signal of resistance degradation and metal-insulator transition.

(8) In accelerated lifetime tests, highly conducting *n*-region surfaces of $SrTiO_3$ behave as a non-blocking cathode and can uptake dissociated atmospheric $O_2$, thus causing local re-oxidation and widespread black electrocoloration by $Fe^{4+}$.

## Appendix: Defect concentration and $PO_2$

Defect calculations usually start with the redox reactions

$$\text{cathodic reaction} \qquad O_o^{\times} \underset{}{\overset{K_{1a}}{\rightleftarrows}} \frac{1}{2}O_2 + V_O^{\bullet\bullet} + 2e' \qquad \text{(A1a)}$$

$$\text{anodic reaction} \qquad \frac{1}{2}O_2 + V_O^{\bullet\bullet} \underset{}{\overset{K_{1b}}{\rightleftarrows}} O_o^{\times} + 2h^{\bullet} \qquad \text{(A1b)}$$

At a fixed concentration of $V_O^{\bullet\bullet}$ $\left(\left[V_O^{\bullet\bullet}\right]\right)$ and Fe $\left(c_A\right)$, however, $PO_2$ and other defect concentrations, $\left[h^{\bullet}\right]$, $\left[e'\right]$, $\left[Fe'_{Ti}\right]$ and $\left[Fe_{Ti}^{x}\right]$ in the Kroger-Vink notation, can be more directly solved from the following equations:

$$Fe_{Ti}^{x} \overset{K_2}{\rightleftarrows} Fe'_{Ti} + h^{\bullet} \qquad \text{(A2)}$$

$$\text{nil} \overset{K_3}{\rightleftarrows} e' + h^{\bullet} \qquad \text{(A3)}$$

$$c_A = \left[Fe_{Ti}^{x}\right] + \left[Fe'_{Ti}\right] \qquad \text{(A4)}$$

$$\left[V_O^{\bullet\bullet}\right] + \left[h^{\bullet}\right] = \left[e'\right] + \left[Fe'_{Ti}\right] \qquad \text{(A5)}$$

Specifically, eliminating $\left[e'\right]$ in favor of $\left[h^{\bullet}\right]$ using Eq. (A3) and eliminating $\left[Fe_{Ti}^{x}\right]$ in favor of $\left[Fe'_{Ti}\right]$ using Eq. (A4) allow Eqs. (A2) and (A5) to be solved. It yields a cubic equation in $\left[h^{\bullet}\right]$ or $\left[Fe'_{Ti}\right]$ that has a unique, analytical solution, from which other concentrations and the local $PO_2$ are obtained from Eqs. (A1), (A3) and (A4), shown in **Figure 1a** and **Figure 5b.** These figures used the thermodynamic and mobility data of Denk *et al.*[28]; in particular, the activation energies for $K_{1a}$, $K_2$ and $K_3$ are, respectively, 4.97 eV, 1.18 eV and 3.3 eV, in addition to entropic contributions. Note that, outside the space charge region, local charge neutrality (Eq. (A5)) is presumably a good approximation.

**References**

[1] Cox GA, Tredgold RH. Time dependence of the electrical conductivity in strontium titanate single crystals. Br J Appl Phys. 1965;16:427-430.

[2] Waser R, Baiatu T, Härdtl K-H. dc Electrical Degradation of Perovskite-Type Titanates: I, Ceramics. J Am Ceram Soc. 1990;73(6):1645-1653.

[3] Waser R, Baiatu T, Härdtl K-H. dc Electrical Degradation of Perovskite-Type Titanates: II, Single Crystals. J Am Ceram Soc. 1990;73(6):1654-1662.

[4] Baiatu T, Waser R, Härdtl K-H. dc Electrical Degradation of Perovskite-Type Titanates: III, A Model of the Mechanism. J Am Ceram Soc. 1990;73(6):1663-1673.

[5] Waser R, Dittmann R, Staikov G, Szot K. Redox-based resistive switching memories—nanoionic mechanisms, prospects, and challenges. Adv Mater. 2009;21:2632-2633.

[6] Waser R, Aono M. Nanoionics-based resistive switching memories. Nat Mater. 2007;6:833-840.

[7] Sawa A. Resistive switching in transition metal oxides. Mater Today. 2008;11:28-36.

[8] Lu Y, Yoon JH, Dong Y, Chen I-W. Purely Electronic Nanometallic Resistance Switching Random-Access Memory. MRS Bulletin. 2018;43:358-64.

[9] Lu Y, Lee JH, Yang X, Chen I-W. Distinguishing Uniform Switching From Filamentary Switching in Resistance Memory Using a Fracture Test. Nanoscale. 2016;8:18113-18120.

[10] Lu Y, Lee JH, Chen I-W. Scalability of Voltage-controlled Filamentary and

Nanometallic Resistance Memory Devices. Nanoscale. 2017;9:12690-12697.

[11] Wang J-J, Huang H-B, Bayer TJM, Moballegh A, Cao Y, Klein A, et al. Defect chemistry and resistance degradation in Fe-doped $SrTiO_3$ single crystal. Acta Mater. 2016;108:229-240.

[12] Havel V, Marchewka A, Menzel S, Hoffmann-Eifert S, Roth G, Waser R. Electroforming of Fe:STO samples for resistive switching made visible by electrocoloration observed by high resolution optical microscopy. Mater Res Soc Symp Proc. 2014;1691.

[13] Wojtyniak M, Szot K, Wrzalik R, Rodenbücher C, Roth G, Waser R. Electro-degradation and resistive switching of Fe-doped $SrTiO_3$ single crystal. J Appl Phys. 2013;113:083713.

[14] Rodewald S, Fleig J, Maier J. Resistance degradation of iron-doped strontium titanate investigated by spatially resolved conductivity measurements. J Am Ceram Soc. 2000;83(8):1969–76.

[15] Rodewald S, Sakai N, Yamaji K, Yokokawa H, Fleig J, Maier J. The effect of the oxygen exchange at electrodes on the high-voltage electrocoloration of Fe-doped $SrTiO_3$ single crystals: a combined SIMS and microelectrode impedance study. J Electroceramics.2001;7:95–105.

[16] Zhao Q-C, Gong H-L, Wang X-H, Chen I-W, Li L-T. Superior Reliability via Two-Step Sintering: Barium Titanate Ceramics. J Amer Ceram Soc. 2016;99(1):191-197.

[17] Yang JJ, Strukov DB, Stewart DR. Memristive devices for computing. Nat Nanotechnol. 2013;8:13-24.

[18] Park J, Chen I-W. In Situ Thermometry Measuring Temperature Flashes Exceeding 1700°C in 8 mol% $Y_2O_3$-Stabilized Zirconia under Constant-Voltage Heating. J Am Ceram Soc. 2013;96(3):697-700.

[19] Dong Y, Chen I-W. Predicting the Onset of Flash Sintering. J Am Ceram Soc. 2015;98(8):2333-2335.

[20] Dong Y, Chen I-W. Onset Criterion for Flash Sintering. J Am Ceram Soc. 2015;98(12):3624-3627.

[21] Jha SK, Raj R. The Effect of Electric Field on Sintering and Electrical Conductivity of Titania. J Am Ceram Soc. 2014;97(2):527-534.

[22] Karakuscu A, Cologna M, Yarotski D, Won J, Francis JSC, Raj R, et al. Defect Structure of Flash-Sintered Strontium Titanate. J Am Ceram Soc. 2012;95(8):2531-2536.

[23] Rheinheimer W, Phuah XL, Wang H, Lemke F, Hoffmann MJ, Wang H. The role of point defects and defect gradients in flash sintering of perovskite oxides. Acta Mater. 2019;165:398-408.

[24] Zhang Y, Nie J, Luo J. Effects of phase and doping on flash sintering of $TiO_2$. J Ceram Soc Japan. 2016;124(4):296–300.

[25] arXiv Preprint

[26] Alvarez A, Dong Y, Chen I-W. DC electrical degradation of YSZ: Voltage-controlled electrical metallization of a fast ion conducting insulator. J Am Ceram Soc. 2020;103:3178-3193.

[27] Waser R. Electronic properties of grain boundaries in $SrTiO_3$ and $BaTiO_3$ ceramics.

Solid State Ionics. 1995;75:89-99.

[28] Denk I, Münch W, Maier J. Partial Conductivities in $SrTiO_3$: Bulk Polarization Experiments, Oxygen Concentration Cell Measurements, and Defect-Chemical Modeling. J Am Ceram Soc.1995;78(12):3265-72.

[29] Langer JB. Instabilities and pattern formation in crystal growth. Rev. Modern Phys. 1980;52(1):1-28.

[30] Witten TA, Sander LM. Diffusion-limited aggregation. Phys Rev B. 1983;27(9):5686-5697.

[31] Janek J, Korte C. Electrochemical blackening of yttria-stabilized zirconia – morphological instability of the moving reaction front. Solid State Ionics. 1999;116:181-195.

[32] Dong Y, Zhang Z, Alvarez A, Chen I-W. Potential jumps at transport bottlenecks cause instability of nominally ionic solid electrolytes in electrochemical cells. Acta Mater. 2020;199:264-277.

[33] Adepalli KK, Yang J, Maier J, Tuller HL, Yildiz B. Tunable Oxygen Diffusion and Electronic Conduction in $SrTiO_3$ by Dislocation-Induced Space Charge Fields. Adv Funct Mater. 2017;27:1700243.

[34] Dong Y, Chen I-W. Oxygen Potential Transition in Mixed Conducting Oxide Electrolyte. Acta Mater. 2018;156(3):399-410.

## Annotated Text

### Annotation A1

**Table 1** Literature review of experimental HALT data

| Material | Thickness (μm) | Voltage (V) | Field (V/cm) | Field exponent *n* | Ref |
|---|---|---|---|---|---|
| $BaTiO_3$-based | – | – | $1.2\times10^4$ | – | 16 |
| $Y^{3+}$-doped $(Ba_{1-x}Ca_xO)_m$-$(Ti_{1-y}Zr_yO_2)$ + MnO + $SiO_2$ | 15 – 30* | – | $1\times10^5$ | 3 | A1 |
| $BaTiO_3$-based | 1.59 | 34 – 44 | – | 2.24 – 2.96 | A2 |
| Mn-doped nano $BaTiO_3$-based | – | – | $1\times10^4$ | – | A3 |
| Dy-doped $BaTiO_3$ | – | – | $1\times10^4$ | – | A4 |
| Mg-doped $BaTiO_3$-based | – | – | $1.2\times10^4$ | – | A5 |
| $BaTiO_3$-based | – | – | $1.2\times10^4$ | – | A6 |
| Ho-doped and Dy-doped $BaTiO_3$-based | – | – | $1.2\times10^4$ | – | A7 |
| BaO-Cao-$SiO_2$ doped $BaTiO_3$-based | 1000 | – | $1\times10^4$ | – | A8 |
| $BaTiO_3$ | – | – | $1\times10^4$ – $1.2\times10^4$ | – | A9 |
| $MnO_2$ and $Y_2O_3$ co-doped $Ba_{0.95}Ca_{0.05}Ti_{0.85}Zr_{0.15}O_3$ | 1000 | – | $1\times10^4$ | – | A10 |
| $Al_2O_3$-coated $BaTiO_3$-based | – | – | $1.35\times10^4$ | – | A11 |
| $BaTiO_3$-based | 1000 | – | $1\times10^4$ | – | A12 |
| $BaTiO_3$-based | – | – | $1.4\times10^5$ – $1.13\times10^6$ | 4.8 – 15.6 | A13 |
| $BaTiO_3$-based | – | 400 | – | – | A14 |
| $MnO_2$ and $Nb_2O$ co-doped 0.9$BaTiO_3$-0.1$(Bi_{0.5}Na_{0.5})TiO_3$ | – | 250 | $1\times10^4$ | – | A15 |
| Type I and Type II dielectric materials | 1500* | 5 – 1500 | – | – | A16 |
| $BaTiO_3$-based | 4.6 | 142 | – | – | A17 |
| $(Ba_{0.95\pm x}Ca_{0.05})(Ti_{0.82}Zr_{0.18})O_3$ | – | 200 | – | – | A18 |
| $(Ba_{1-x}Ca_x)_z(Ti_{0.99-y}Zr_yMn_{0.01})O_3$ | 400 – 450 | 200 | – | – | A19 |
| $Ca(Zr_{1-x}Ti_x)O_3$ | – | 800 – 1000 | $5\times10^5$ – $6.25\times10^5$ | 2.14 | A20 |

| **Material** | **Thickness (μm)** | **Voltage (V)** | **Field (V/cm)** | **Field exponent *n*** | **Ref** |
|---|---|---|---|---|---|
| Mn-doped $BaTiO_3$ | 3.5 | – | $5\times10^4$–$2\times10^5$ | – | A21 |
| $Ba(Ti,Zr)O_3$-based | 2* | – | $1\times10^5$ | – | A22 |
| $BaTiO_3$-based | 10* | – | $1\times10^5$ | | A23 |
| $BaTiO_3$-based | 12 | 400 | – | – | A24 |
| Y-doped $BaTiO_3$ | – | 400 – 600 | – | 3 – 5 | A25 |
| Commercial MLCC | 36.5 | 75 – 200 | – | 2.46 | A26 |
| Commercial Base Metal Electrode MLCC | ~8 | 250 – 600 | – | 4.5 – 5.5 | A27 |
| PZT | 0.88 | – | $2.84\times10^5$ – $4.55\times10^5$ | 5.3 | A28 |
| PZT | 1.5 | – | $2\times10^5$ – $3\times10^5$ | 4.24 | A29 |
| Nb-doped PZT | 0.8 | 8 – 45 | – | 4 – 5 | A30 |

*Green thickness

**Annotation A2**

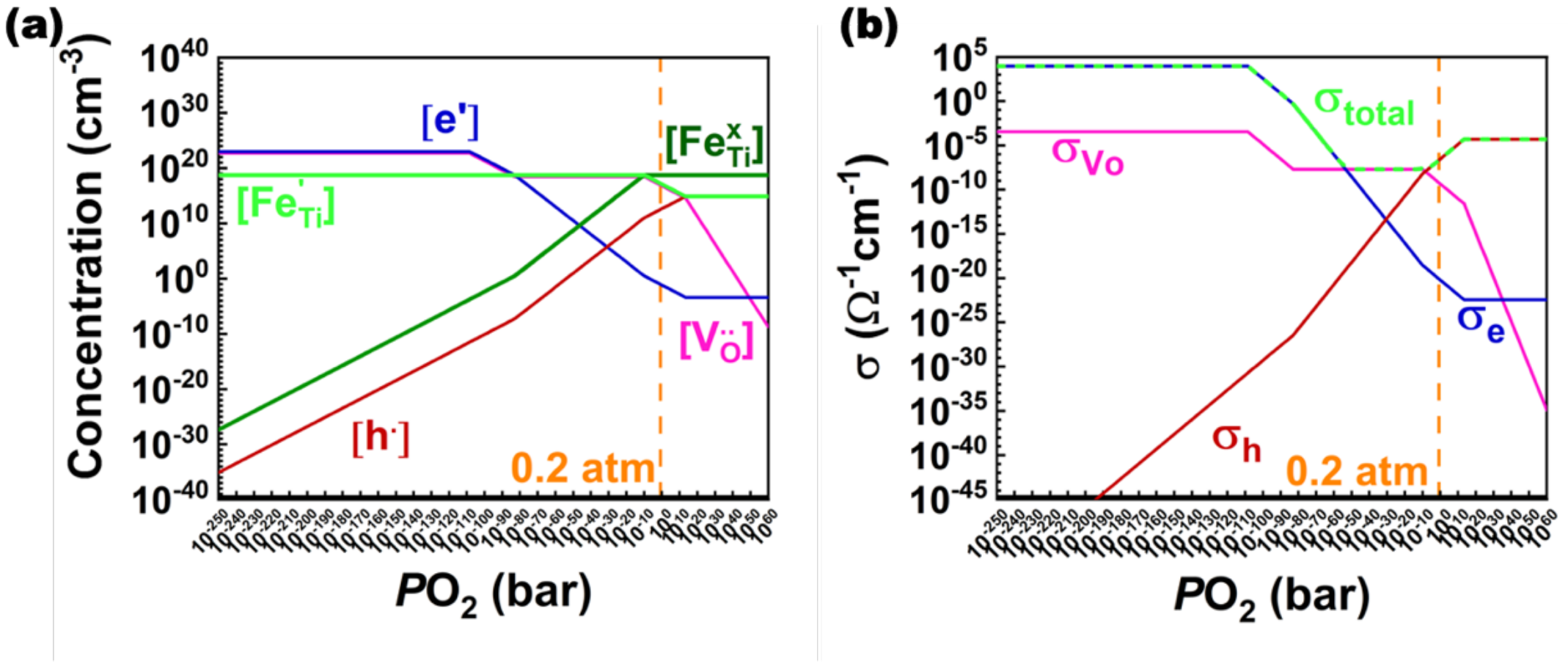

**Figure A1** Calculated (a) defect concentrations ($[Fe'_{Ti}]$, $[Fe^{x}_{Ti}]$, $[h^{\cdot}]$, $[e']$, and $[V_O^{\cdot\cdot}]$) and (b) $\sigma_{V_O}$, $\sigma_e$ and $\sigma_h$ plus their sum at 210°C with $5.58\times10^{18}$ cm$^{-3}$ Fe concentration.

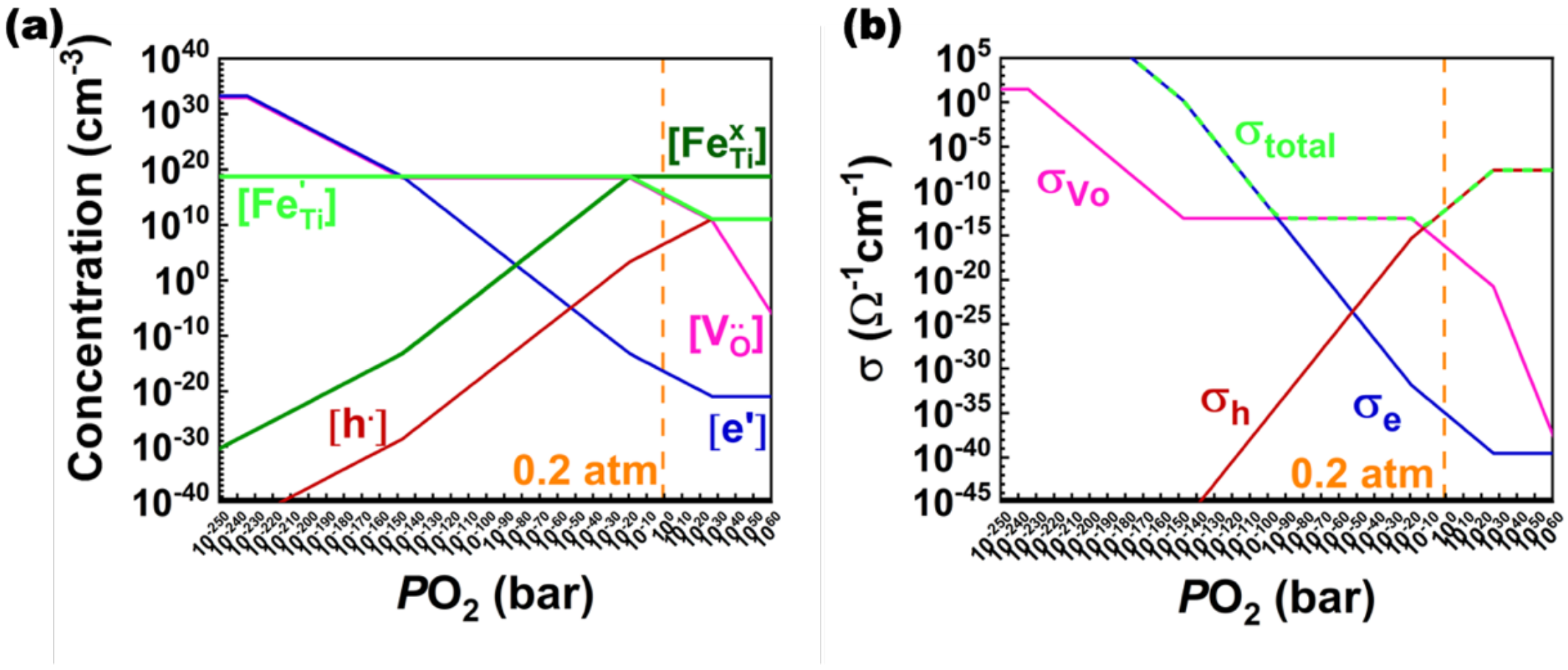

**Figure A2** Calculated (a) defect concentrations ($\left[\mathrm{Fe}'_{\mathrm{Ti}}\right]$, $\left[\mathrm{Fe}^{\mathrm{x}}_{\mathrm{Ti}}\right]$, $\left[\mathrm{h}^{\bullet}\right]$, $\left[\mathrm{e}'\right]$, and $\left[\mathrm{V}_{\mathrm{O}}^{\bullet\bullet}\right]$) and (b) $\sigma_{\mathrm{V_O}}$, $\sigma_{\mathrm{e}}$ and $\sigma_{\mathrm{h}}$ plus their sum at 25°C with $5.58\times10^{18}$ $\mathrm{cm}^{-3}$ Fe concentration.

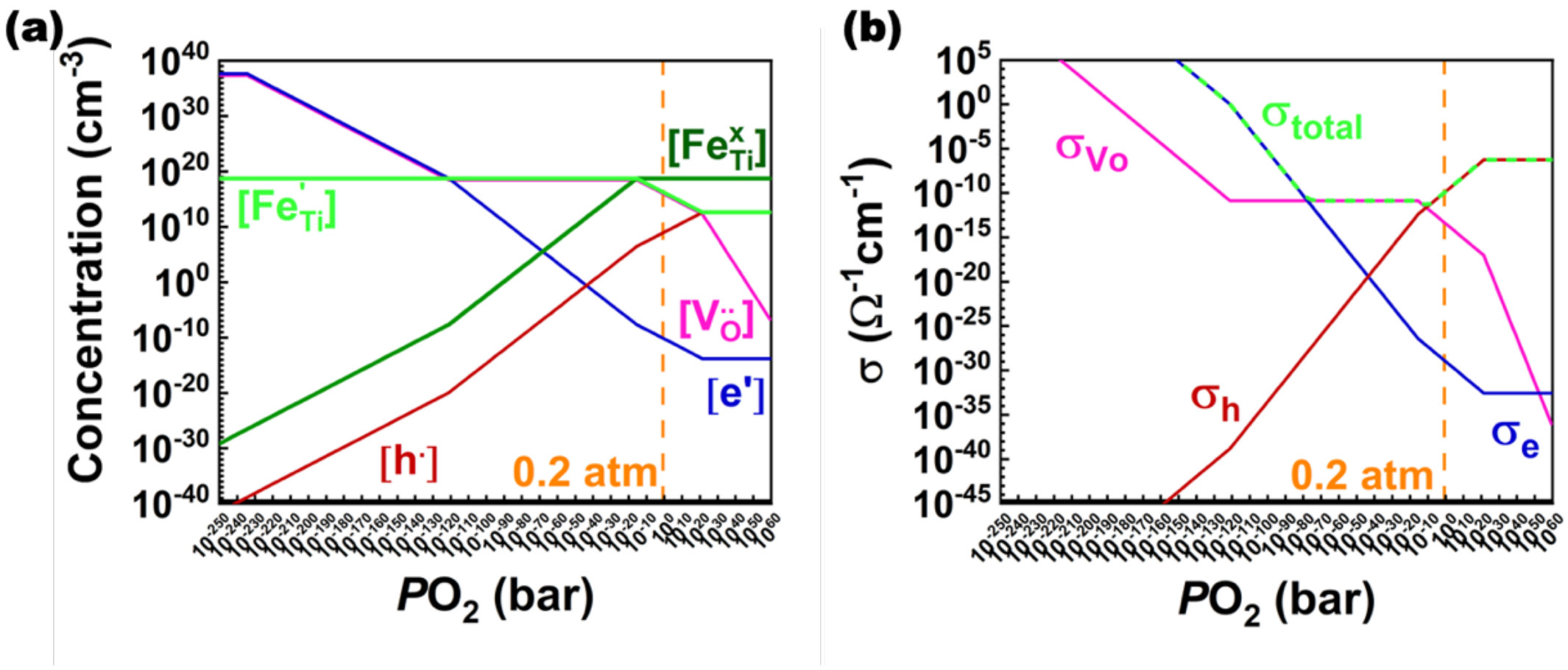

**Figure A3** Calculated (a) defect concentrations ($\left[\mathrm{Fe}'_{\mathrm{Ti}}\right]$, $\left[\mathrm{Fe}^{\mathrm{x}}_{\mathrm{Ti}}\right]$, $\left[\mathrm{h}^{\bullet}\right]$, $\left[\mathrm{e}'\right]$, and $\left[\mathrm{V}_{\mathrm{O}}^{\bullet\bullet}\right]$) and (b) $\sigma_{\mathrm{V_O}}$, $\sigma_{\mathrm{e}}$ and $\sigma_{\mathrm{h}}$ plus their sum at 80°C with $5.58\times10^{18}$ $\mathrm{cm}^{-3}$ Fe concentration.

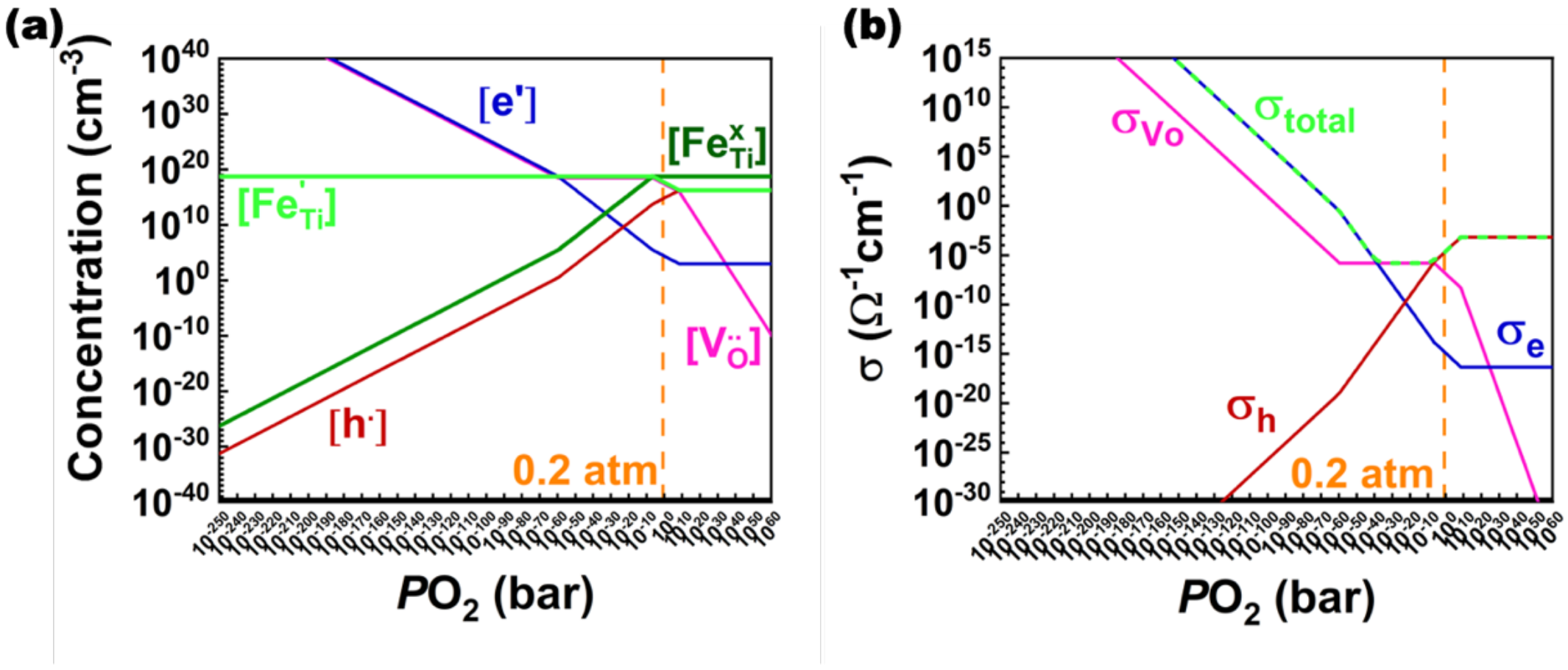

**Figure A4** Calculated (a) defect concentrations ($\left[\mathrm{Fe}'_{\mathrm{Ti}}\right]$, $\left[\mathrm{Fe}^{\mathrm{x}}_{\mathrm{Ti}}\right]$, $\left[\mathrm{h}^{\cdot}\right]$, $\left[\mathrm{e}'\right]$, and $\left[\mathrm{V}_{\mathrm{O}}^{\cdot\cdot}\right]$) and (b) $\sigma_{\mathrm{V_O}}$, $\sigma_{\mathrm{e}}$ and $\sigma_{\mathrm{h}}$ plus their sum at 350°C with $5.58\times10^{18}$ cm$^{-3}$ Fe concentration.

## Annotation A3

Ignoring the potential drop in the *n*-region and denoting the potential drops in the *p*- and *i*-regions as $\Delta\phi_1$ and $\Delta\phi_2$, respectively, we have

$$\Delta\phi = \Delta\phi_1 + \Delta\phi_2 = \phi_{\mathrm{A}} - \phi_{\mathrm{C}} \tag{A6}$$

Denoting particle (d = $\mathrm{h}^{\cdot}$, $\mathrm{e}'$, or $\mathrm{V}_{\mathrm{O}}^{\cdot\cdot}$) charge ($ze$) by $z_{\mathrm{d}}e$, particles flux ($j$) by $j_{\mathrm{d}}$, and their electric flux ($J$) by $J_{\mathrm{d}}$, we impose continuity of electrical current across all three regions. Across the interface at $x_1$ where the hole current passes over to the oxygen-vacancy current,

$$J_{\mathrm{h}} = z_{\mathrm{h}} e j_{\mathrm{h}} = z_{\mathrm{V_O}} e j_{\mathrm{V_O}} = J_{\mathrm{V_O}} \tag{A7}$$

Denoting particle mobility ($M$) and diffusivity ($D$) by $M_{\mathrm{d}} = D_{\mathrm{d}}/kT$, we obtain

$$j_{\mathrm{h}} = z_{\mathrm{h}} e \left(M_{\mathrm{h}}/L\right)\left[\mathrm{h}^{\cdot}\right]^{\mathrm{f}} \Delta\phi_1/x_1 \tag{A8a}$$

$$j_{\mathrm{V_O}} = z_{\mathrm{V_O}} e \left(M_{\mathrm{V_O}}/L\right)\left[\mathrm{V}_{\mathrm{O}}^{\cdot\cdot}\right]^{\mathrm{i}} \Delta\phi_2/\left(x_2 - x_1\right) \tag{A8b}$$

throughout the $p$- and $i$-region. Together, Eq. (A7-A8) give

$$z_{\mathrm{h}}^{2}\left(M_{\mathrm{h}}/L\right)\left[\mathrm{h}^{\bullet}\right]^{\mathrm{f}}\Delta\phi_{1}/x_{1}=z_{\mathrm{V_O}}^{2}\left(M_{\mathrm{V_O}}/L\right)\left[\mathrm{V_O^{\bullet\bullet}}\right]^{\mathrm{i}}\Delta\phi_{2}/\left(x_{2}-x_{1}\right) \tag{A9}$$

Combining Eq. (A6, A9) gives the following results which are just the Ohm's law and Kirchhoff's law for a cross section of unit area

$$R_{1}=\frac{L_{1}}{\sigma_{1}}=\frac{x_{1}L}{z_{\mathrm{h}}^{2}e^{2}M_{\mathrm{h}}\left[\mathrm{h}^{\bullet}\right]^{\mathrm{f}}} \tag{A10a}$$

$$R_{2}=\frac{L_{2}}{\sigma_{1}}=\frac{\left(x_{2}-x_{1}\right)L}{z_{\mathrm{V_O}}^{2}e^{2}M_{\mathrm{V_O}}\left[\mathrm{V_O^{\bullet\bullet}}\right]^{\mathrm{i}}} \tag{A10b}$$

$$\frac{\Delta\phi_{1}}{\Delta\phi_{2}}=\frac{R_{1}}{R_{1}+R_{2}}=\frac{\dfrac{z_{\mathrm{V_O}}^{2}M_{\mathrm{V_O}}\left[\mathrm{V_O^{\bullet\bullet}}\right]^{\mathrm{i}}}{x_{2}-x_{1}}}{\dfrac{z_{\mathrm{h}}^{2}M_{\mathrm{h}}\left[\mathrm{h}^{\bullet}\right]^{\mathrm{f}}}{x_{1}}+\dfrac{z_{\mathrm{V_O}}^{2}M_{\mathrm{V_O}}\left[\mathrm{V_O^{\bullet\bullet}}\right]^{\mathrm{i}}}{x_{2}-x_{1}}} \tag{A10c}$$

$$\frac{\Delta\phi_{2}}{\Delta\phi}=\frac{R_{2}}{R_{1}+R_{2}}=\frac{\dfrac{z_{\mathrm{h}}^{2}M_{\mathrm{h}}\left[\mathrm{h}^{\bullet}\right]^{\mathrm{f}}}{x_{1}}}{\dfrac{z_{\mathrm{h}}^{2}M_{\mathrm{h}}\left[\mathrm{h}^{\bullet}\right]^{\mathrm{f}}}{x_{1}}+\dfrac{z_{\mathrm{V_O}}^{2}M_{\mathrm{V_O}}\left[\mathrm{V_O^{\bullet\bullet}}\right]^{\mathrm{i}}}{x_{2}-x_{1}}} \tag{A10d}$$

If the potential drop in the $n$-type region $\Delta\phi_{3}$ is desired, it can be obtained by requiring, at $x_{2}$, $z_{\mathrm{V_O}}ej_{\mathrm{V_O}}=J_{\mathrm{V_O}}=J_{\mathrm{e}}$, with $j_{\mathrm{e}}=z_{\mathrm{e}}e\left(M_{\mathrm{e}}/L\right)\left[\mathrm{e}'\right]^{\mathrm{f}}\Delta\phi_{3}/\left(1-x_{2}\right)$.

The electric flux is carried by the particle flux, and since we have assumed box profiles of defect concentrations that are discontinuous across the interfaces of different regions, a continuous particle flux will cause interface to move. Specifically, the $i$-region will shrink and the other two regions will grow. Mathematically, $\left[\mathrm{V_O^{\bullet\bullet}}\right]$ conservation at the moving interfaces $x_{1}$ and $x_{2}$ gives

$$\frac{dx_{1}}{dt}=\frac{j_{\mathrm{V_O}}}{L\left[\mathrm{V_O^{\bullet\bullet}}\right]^{\mathrm{i}}} \tag{A11}$$

$$\frac{dx_2}{dt} = -\frac{j_{\mathrm{V_O}}}{L\left(\left[\mathrm{V_O^{\bullet\bullet}}\right]^{\mathrm{f}} - \left[\mathrm{V_O^{\bullet\bullet}}\right]^{\mathrm{i}}\right)} \tag{A12}$$

These two velocities are related to each other, giving

$$dx_2 = -\left(\frac{\left[\mathrm{V_O^{\bullet\bullet}}\right]^{\mathrm{i}}}{\left[\mathrm{V_O^{\bullet\bullet}}\right]^{\mathrm{f}} - \left[\mathrm{V_O^{\bullet\bullet}}\right]^{\mathrm{i}}}\right) dx_1 \tag{A13}$$

which with the initial condition of $x_1 = 0$ and $x_2 = 1$ gives the following solution

$$x_2 = 1 - \left(\frac{\left[\mathrm{V_O^{\bullet\bullet}}\right]^{\mathrm{i}}}{\left[\mathrm{V_O^{\bullet\bullet}}\right]^{\mathrm{f}} - \left[\mathrm{V_O^{\bullet\bullet}}\right]^{\mathrm{i}}}\right) x_1 \tag{A14}$$

Therefor, $x_1$ and $x_2$ are constrained to move in a straight line in the ($x_1, x_2$) plane.

The interface movement ends when $x_1 = x_2 = x^{\mathrm{f}}$, when the *i*-region is pinched off. At this point, Eq. (A14) reduces to

$$\left[\mathrm{V_O^{\bullet\bullet}}\right]^{\mathrm{i}} = \left(1 - x^{\mathrm{f}}\right)\left[\mathrm{V_O^{\bullet\bullet}}\right]^{\mathrm{f}} \quad \text{or} \quad x^{\mathrm{f}} = \frac{\left[\mathrm{V_O^{\bullet\bullet}}\right]^{\mathrm{f}} - \left[\mathrm{V_O^{\bullet\bullet}}\right]^{\mathrm{i}}}{\left[\mathrm{V_O^{\bullet\bullet}}\right]^{\mathrm{f}}} = 1 - \frac{\left[\mathrm{V_O^{\bullet\bullet}}\right]^{\mathrm{i}}}{\left[\mathrm{V_O^{\bullet\bullet}}\right]^{\mathrm{f}}} \tag{A15}$$

which is just a statement of $\left[\mathrm{V_O^{\bullet\bullet}}\right]$ conservation. From then on, the steady state with only the electronic (e and h) current left takes over.

It is now clear that all the potentials are known once $x_1$ and $x_2$ are, and since $x_1$ and $x_2$ are constrained to move in a straight line, only one of them need to be solved as a function of time. This can be done by integrating Eq. (A11) or Eq. (A12). As shown below, the solution can be cast into a parabolic form with a time constant that scales with $L^2/2D_{\mathrm{V_O}}$.

Specifically, combining Eq. (A8b), (A10d) and (A11), we obtain

$$\frac{dx_1}{dt} = \frac{z_{\mathrm{V_O}} e\left(\frac{M_{\mathrm{V_O}}}{L^2}\right)\left(\frac{\Delta\phi}{x_2 - x_1}\right)\left(\frac{z_{\mathrm{h}}^2 M_{\mathrm{h}}\left[\mathrm{h}^{\bullet}\right]^{\mathrm{f}}}{x_1}\right)}{\frac{z_{\mathrm{h}}^2 M_{\mathrm{h}}\left[\mathrm{h}^{\bullet}\right]^{\mathrm{f}}}{x_1} + \frac{z_{\mathrm{V_O}}^2 M_{\mathrm{V_O}}\left[\mathrm{V_O^{\bullet\bullet}}\right]^{\mathrm{i}}}{x_2 - x_1}}$$

$$= \frac{z_{\mathrm{V_O}} e\left(\frac{M_{\mathrm{V_O}}}{L^2}\right)\left(\frac{\Delta\phi}{x_2 - x_1}\right)}{1 + \left(\frac{x_1}{x_2 - x_1}\right)\left(\frac{z_{\mathrm{V_O}}^2 M_{\mathrm{V_O}}\left[\mathrm{V_O^{\bullet\bullet}}\right]^{\mathrm{i}}}{z_{\mathrm{h}}^2 M_{\mathrm{h}}\left[\mathrm{h}^{\bullet}\right]^{\mathrm{f}}}\right)}$$

$$= \frac{z_{\mathrm{V_O}} e\Delta\phi\left(\frac{M_{\mathrm{V_O}}}{L^2}\right)}{x_2 - x_1 + \left(\frac{z_{\mathrm{V_O}}^2 M_{\mathrm{V_O}}\left[\mathrm{V_O^{\bullet\bullet}}\right]^{\mathrm{i}}}{z_{\mathrm{h}}^2 M_{\mathrm{h}}\left[\mathrm{h}^{\bullet}\right]^{\mathrm{f}}}\right) x_1}$$

Using Eq. (A14-A15), we further simplify the above into

$$\frac{dx_1}{dt} = \frac{z_{\mathrm{V_O}} e\Delta\phi\left(\frac{M_{\mathrm{V_O}}}{L^2}\right)}{1 - (1-\gamma)\left(\frac{x_1}{x^{\mathrm{f}}}\right)} \tag{A16}$$

with

$$\gamma = \left(\frac{z_{\mathrm{V_O}}^2 M_{\mathrm{V_O}}\left[\mathrm{V_O^{\bullet\bullet}}\right]^{\mathrm{i}}}{z_{\mathrm{h}}^2 M_{\mathrm{h}}\left[\mathrm{h}^{\bullet}\right]^{\mathrm{f}}}\right) x^{\mathrm{f}} = \left(\frac{z_{\mathrm{V_O}}^2 M_{\mathrm{V_O}}\left[\mathrm{V_O^{\bullet\bullet}}\right]^{\mathrm{i}}}{z_{\mathrm{h}}^2 M_{\mathrm{h}}\left[\mathrm{h}^{\bullet}\right]^{\mathrm{f}}}\right)\left(1 - \frac{\left[\mathrm{V_O^{\bullet\bullet}}\right]^{\mathrm{i}}}{\left[\mathrm{V_O^{\bullet\bullet}}\right]^{\mathrm{f}}}\right) \tag{A17}$$

Writing $M_{\mathrm{d}}$ as $D_{\mathrm{d}}/kT$, Eq. (A17) can be integrated to obtain

$$\frac{x_1}{x^{\mathrm{f}}} = \frac{1}{1-\gamma}\left(1 - \sqrt{1 - \frac{t}{t^{\mathrm{f}}} + \gamma^2\left(\frac{t}{t^{\mathrm{f}}}\right)}\right) \tag{A18}$$

with $t^{\mathrm{f}}$ being the time reaching $x_1 = x_2 = x^{\mathrm{f}}$,

$$\frac{t^{\mathrm{f}}}{L^2 / 2D_{\mathrm{V_O}}} = (1+\gamma)\left(\frac{kT}{z_{\mathrm{V_O}} e\Delta\phi}\right)\left(1 - \frac{\left[\mathrm{V_O^{\bullet\bullet}}\right]^{\mathrm{i}}}{\left[\mathrm{V_O^{\bullet\bullet}}\right]^{\mathrm{f}}}\right) = (1+\gamma)\left(\frac{kT}{z_{\mathrm{V_O}} e\Delta\phi}\right) x^{\mathrm{f}} \tag{A19}$$

Finally, the electric flux $J$ being the same throughout the sample can be obtained from $J$ in any region, say the *i*-region, which is $J = J_{V_O} = z_{V_O} e j_{V_O}$ or

$$J = z_{V_O}^2 e^2 \left[V_O^{\bullet\bullet}\right]^{i} \left(\frac{M_{V_O}}{L}\right)\left(\frac{\Delta\phi_2}{x_2 - x_1}\right)$$

$$= z_{V_O}^2 e^2 \left[V_O^{\bullet\bullet}\right]^{i} \left(\frac{M_{V_O}}{L}\right)\left(\frac{\Delta\phi}{1-(1-\gamma)\left(\frac{x_1}{x^{f}}\right)}\right)$$

$$= z_{V_O}^2 e^2 M_{V_O} \left[V_O^{\bullet\bullet}\right]^{i} \left(\frac{\Delta\phi}{L}\right)\left(\frac{1}{\sqrt{1-\frac{t}{t^{f}}+\gamma^2\left(\frac{t}{t^{f}}\right)}}\right)$$

$$= \frac{\sigma_{V_O} E}{\sqrt{1-\frac{t}{t^{f}}+\gamma^2\left(\frac{t}{t^{f}}\right)}} \qquad \text{(A20)}$$

where in the last equality we used the ion conductivity $\sigma_{V_O} = z_{V_O}^2 e^2 \left[V_O^{\bullet\bullet}\right]^{i} M_{V_O}$ and the nominal electric field $E = \Delta\phi / L$.

It is easy to verify that the two limiting currents are, $J^{i}$ at $t = 0$ and $J^{f}$ at $t = t^{f}$, after that the current no longer changes,

$$J(t=0) = J^{i} = \sigma_{V_O} E \quad \text{and} \quad J\left(t = t^{f}\right) = J_{\text{steady state}} = \frac{\sigma_{V_O} E}{\gamma} = \frac{\sigma_{h} E \left[V_O^{\bullet\bullet}\right]^{f}}{\left[V_O^{\bullet\bullet}\right]^{f} - \left[V_O^{\bullet\bullet}\right]^{i}} \qquad \text{(A21)}$$

where $\sigma_h = z_h^2 e^2 \left[h^{\bullet}\right]^{i} M_h$ is the hole conductivity. Thus, $\gamma$ can also be expressed as

$$\gamma = J^{i} / J^{f} \qquad \text{(A22)}$$

The solution is now complete.

**Annotation A4**

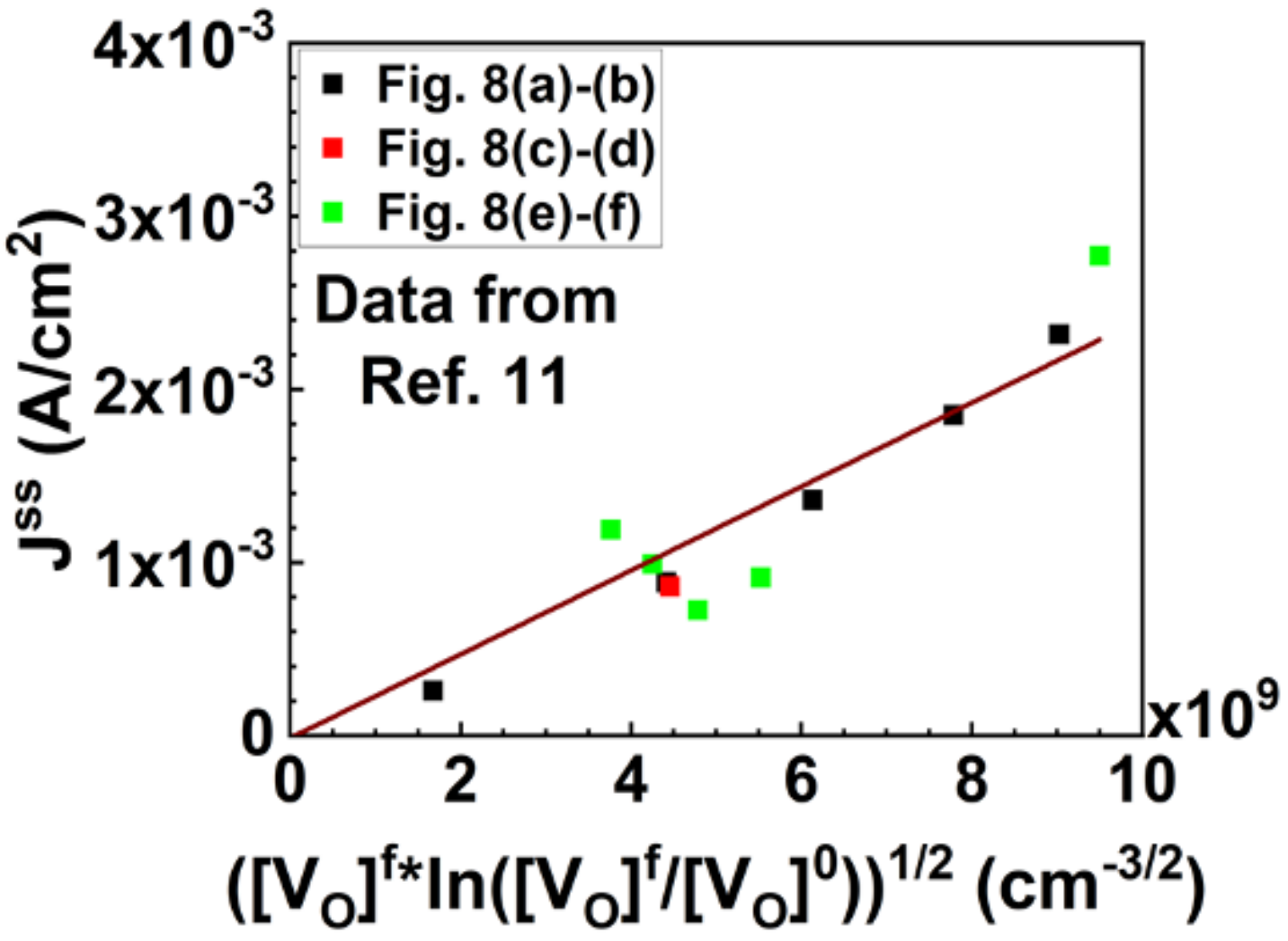

**Figure A5** Numerical solutions of steady state conductivity vs concentrations from Fig. 8 in Ref. 11 follow linear fit motivated by Eq. (12d).

**Annotation A5**

Waser *et al.*[2-3] reported more variance in undoped samples than acceptor-doped ones, which we did not see.

**Annotation A6**

After entering compliance control, there is a slight decrease of electroluminescence over time. This is because as the number of electrons gradually increases in this stage (due to oxygen loss, which will become clear later from observations of oxygen bubbles), the voltage to maintain the same $J$ decreases (this is why the voltage at the beginning of cooling in **Figure 5a** is much less than the voltage used before the compliance limit was hit), which leads to less electron acceleration, hence less electroluminescence.

## References

[A1] Sato S, Nakano Y, Sato A, Nomura T. Effect of Y-Doping on Resistance Degradation of Multilayer Ceramic Capacitors with Ni Electrodes under the Highly Accelerated Life Test. Jpn J Appl Phys. 1997;36:6016-20.

[A2] Yoon J-R, Lee K-M, Lee S-W. Analysis the Reliability of Multilayer Ceramic Capacitor with inner Ni Electrode under highly Accelerated Life Test Conditions. Trans Electr Electron Mater. 2009;10(1):5-8.

[A3] Gong H, Wang X, Zhang S, Tian Z, Li L. Electrical and reliability characteristics of Mn-doped nano $BaTiO_3$-based ceramics for ultrathin multilayer ceramic capacitor application. J Appl Phys. 2012;112:114119.

[A4] Gong H, Wang X, Zhang S, Yang X, Li L. Influence of sintering temperature on core-shell structure evolution and reliability in Dy modified $BaTiO_3$ dielectric ceramics. Phys Status Solidi A. 2014;5:1213-8.

[A5] Gong H, Wang X, Zhao Q, Li L. Effect of Mg on the dielectric and electrical properties of $BaTiO_3$-based ceramics. J Mater Sci. 2015;50:6808-6906.

[A6] Gong H, Wang X, Zhang S, Li L. Sintering behavior and reliability characteristics of $BaTiO_3$-based ceramics prepared by different methods. J Mater Sci. 2015;50:3523-30.

[A7] Gong H, Wang X, Zhang S, Li L. Synergistic effect of rare-earth elements on the dielectric properties and reliability of $BaTiO_3$-based ceramics for multilayer ceramic capacitors. Mater Res Bull. 2016;73:233-9.

[A8] Zhao Q, Gong H, Wang X, Luo B, Li L. Influence of $BaO$-$CaO$-$SiO_2$ on dielectric

properties and reliability of $BaTiO_3$-based ceramics. Phys Status Solidi A. 2016;213(4):1077-81.

[A9] Gong H, Wang X, Zhang S, Wen H, Li L. Grain size effect on electrical and reliability characteristics of modified fine-grained $BaTiO_3$ ceramics for MLCCs. J Euro Ceram Soc. 2014;34:1733-9.

[A10] Shen Z, Wang X, Gong H, Wu L, Li L. Effect of $MnO_2$ on the electrical and dielectric properties of Y-doped $Ba_{0.95}Ca_{0.05}Ti_{0.85}Zr_{0.15}O_3$ ceramics in reducing atmosphere. Ceram Int 2014;40:13833-9.

[A11] Zhao Q, Wang X, Gong H, Liu B, Luo B, Li L. The properties of $Al_2O_3$ coated fine-grain temperature stable $BaTiO_3$-based ceramics sintered in reducing atmosphere. J Am Ceram Soc. 2018;101:1245-54.

[A12] Zhu C, Zhao Q, Cai Z, Guo L, Li L, Wang X. High reliable non-reducible non-reducible ultra-fine $BaTiO_3$-based ceramics fabricated via solid-state method. J Alloys Compd. 2020;829:154496

[A13] Randall CA, Maier R, Qu W, Kobayashi K, Morita K, Mizuno Y, et al. Improved reliability predictions in high permittivity dielectric oxide capacitors under high dc electrics fields with oxygen vacancy induced electromigration. J Appl Phys. 2013;113:014101.

[A14] Munikoti R, Dhar P. Highly Accelerated Life Testing (HALT) for Multilayer Ceramic Capacitor Qualification. IEEE Trans Components, Hybrids Manuf Technol. 1988;11(4):342-5.

[A15] Chen L, Wang H, Zhao P, Shen Z, Zhu C, Cen Z, Li L, Wang X. Effect of $MnO_2$

on the dielectric properties of Nb-doped $BaTiO_3$-$(Bi_{0.5}Na_{0.5})TiO_3$ ceramics for X9R MLCC applications. J Am Ceram Soc. 2019;102;2781-2790.

[A16] Bernard J, Houivet D, El fallah J, Haussonne J-M. Effect of hygrometry on dielectric materials. J Euro Ceram Soc. 2004;24:1509-11.

[A17] Dale G, Strawhorne M, Sinclair DC, Dean JS. Finite element modeling on the effect of intra-granular porosity on the dielectric properties of $BaTiO_3$ MLCCs. J Am Ceram Soc. 2018;101:1211-20.

[A18] Jain TA, Fung KZ, Chan J. Effect of the A/B ratio on the microstructures and electrical properties of $(Ba_{0.95\pm x}Ca_{0.05})(Ti_{0.82}ZrO_{0.18})O_3$ for multilayer ceramic capacitors with nickel electrodes. J Alloys Compd. 2009;468:370-4.

[A19] Lee W-H, Tseng T-Y, Hennings D. Effects of A/B cation ratio on the microstructure and lifetime of $(Ba_{1-x}Ca_x)_z(Ti_{0.99-y}Zr_yMn_{0.01})O_3$ (BCTZM) sintered in reducing atmosphere. J Mater Sci: Mater Electron. 2000;11:157-62.

[A20] Lee H, Kim JR, Lanagan MJ, Trolier-McKinstry S, Randall CA. High-Energy Density Dielectrics and Capacitors for Elevated Temperatues: $Ca(Zr,Ti)O_3$. J Am Ceram Soc. 2013;96(4):1209-13.

[A21] Morita K, Mizuno Y, Chazono H, Kishi H. Effect of Mn Addition on dc-Electrical Degradation of Multilayer Ceramic Capacitor with Ni Internal Electrode. Jpn J Appl Phys. 2002;41:6975-61.

[A22] Okamatsu T, Sano H, Takagi H. Effects of Grain Boundary and Segregated-Phases on Reliability of $Ba(Ti,Zr)O_3$-Based Ni Electrode MLCs. Jpn J Appl Phys. 2005;44:7186-9.

[A23] Sakabe Y, Hamaji Y, Sano H, Wada N. Effects of Rare-Earth Oxides on the Reliability of X7R Dielectrics. Jpn J Appl Phys. 2002;41:5668-73.

[A24] Yang GY, Lian GD, Dickey EC, Randall CA, Barber DE, et al. Oxygen nonstoichiometry and dielectric evolution of $BaTiO_3$. Part II-insulation resistance degradation under applied dc bias. J Appl Phys 2004;96:7500.

[A25] Hernández-López AM, Aguilar-Garib JA, Guillemet-Fritsch S, Nava-Quintero R, Dufour P, Tenailleau C, et al. Reliability of X7R Multilayer Ceramic Capacitors During High Accelerated Life Testing (HALT). Mater. 2018;11,1900.

[A26] Minford WJ. Accelerated Life Testing and Reliability of High K Multilayer Ceramic Capacitors. IEEE Trans Components, Hybrids Manuf Technol. 1982;5:297-300.

[A27] Paulsen JL, Reed EK. Highly accelerated lifetesting of base-metal-electrode ceramic chip capacitors. Microelectron Reliab. 2002;42:815-20.

[A28] Polcawich RG, Feng C, Vanatta P, Piekarz R, Trolier-McKinstry S, Dubey M, et al. Highly Accelerated Lifetime Testing (HALT) of Lead Zirconate Titanate (PZT) Thin Films. Proc 12th IEEE Inter Symp Appl Ferro 2001;1:357-60.

[A29] Ko SW, Zhu W, Fragkiadakis C, Borman T, Wang K, Mardilovich P, et al. Improvement of reliability and dielectric breakdown strength of Nb-doped lead zirconate films via microstructure control of seed. J Am Ceram Soc. 2019;102:1211-7.

[A30] Al-Shareef H, Dimos D. Accelerated Life-Time Testing and Resistance Degradation of Thin-Film Decoupling Capacitors. Proc Tenth IEEE Int Symp Appl Ferroelectr. 1996;1:421-5.